# On Semantic Generalizations of the Bernays-Schönfinkel-Ramsey Class with Finite or Co-finite Spectra

Abhisekh Sankaran and Supratik Chakraborty
Indian Institute of Technology (IIT), Bombay, India
{abhisekh, supratik}@cse.iitb.ac.in

## Abstract

*Motivated by model-theoretic properties of the Bernays-Schönfinkel-Ramsey (BSR) class, we present a family of semantic classes of FO formulae with finite or co-finite spectra over a relational vocabulary $\Sigma$. A class in this family is denoted $\mathbf{EBS}_\Sigma(\sigma)$, where $\sigma \subseteq \Sigma$. Formulae in $\mathbf{EBS}_\Sigma(\sigma)$ are preserved under substructures modulo a bounded core and modulo re-interpretation of predicates in $\Sigma \setminus \sigma$. We study several properties of the family $\mathcal{EBS}_\Sigma = \{\mathbf{EBS}_\Sigma(\sigma) \mid \sigma \subseteq \Sigma\}$. For example, classes in $\mathcal{EBS}_\Sigma$ are spectrally indistinguishable, the smallest class, $\mathbf{EBS}_\Sigma(\Sigma)$, is semantically equivalent to BSR over $\Sigma$, and the largest class, $\mathbf{EBS}_\Sigma(\emptyset)$, is the set of all FO formulae over $\Sigma$ with finite or co-finite spectra. Furthermore, $(\mathcal{EBS}_\Sigma, \subseteq)$ forms a lattice that is isomorphic to the powerset lattice $(\wp(\Sigma), \subseteq)$. We also show that if $\Sigma$ contains at least one predicate of arity $\geq 2$, there exist semantic gaps between $\mathbf{EBS}_\Sigma(\sigma_1)$ and $\mathbf{EBS}_\Sigma(\sigma_2)$ if $\sigma_1 \neq \sigma_2$. This gives a natural semantic generalization of BSR as ascending chains in the lattice $(\mathcal{EBS}_\Sigma, \subseteq)$.*

*We study the semantic relationship of the $\mathbf{EBS}_\Sigma(\sigma)$ classes with other fragments of FO, and observe that many well-known classes are semantically subsumed by $\mathbf{EBS}_\Sigma(\Sigma)$ or $\mathbf{EBS}_\Sigma(\emptyset)$. This study also provides alternative proofs of some interesting results like the Loś-Tarski Theorem and the semantic subsumption of the Löwenheim class with equality by BSR.*

*We show that the membership problem for each class in $\mathcal{EBS}_\Sigma$ is undecidable. As a step towards syntactically characterizing fragments of $\mathbf{EBS}_\Sigma(\sigma)$, we present a syntactic sub-class of $\mathbf{EBS}_\Sigma(\sigma)$ called $\mathbf{EDP}_\Sigma(\sigma)$ and give an expression for the size of the bounded cores of models of $\mathbf{EDP}_\Sigma(\sigma)$ formulae. Using this, we characterize the complexity of the satisfiability problem for $\mathbf{EDP}_\Sigma(\sigma)$. We show that the $\mathbf{EDP}_\Sigma(\sigma)$ classes also form a lattice structure. Finally, we study some closure properties and applications of the classes presented.*

## 1 Introduction and Preliminaries

The Bernays-Schönfinkel-Ramsey class (henceforth called **BSR**) is a widely used decidable class of first order logic sentences [1]. Several recent works have addressed the problem of efficiently checking satisfiability of **BSR** formulae in practice [3]. Motivated by the increasing interest in **BSR** formulae in recent times, we consider the problem of generalizing this class both in a semantic and syntactic sense, while retaining decidability of satisfiability checking. In doing so, we would like to exploit advances in decision procedures for **BSR** to answer corresponding questions for formulae in these more general classes

***Preliminaries:*** We represent a tuple of variables $(x_1, x_2, \ldots x_n)$ using the shorthand $\mathbf{x}$. A first order logic (henceforth called **FO**) formula $\varphi(x_1, x_2, \ldots x_n)$ with *free variables* $x_1, x_2, \ldots x_n$ is represented as $\varphi(\mathbf{x})$ for notational convenience. A *sentence* is a formula without any free variables. The *signature* or *vocabulary* of a formula is the set of all function and predicate symbols that appear in the formula. It is well known that functions are "syntactic sugar" in first-order logic. Specifically, for every formula $\varphi$ with a $k$-ary function $f$, there is an equisatisfiable formula $\psi$ with a $(k+1)$-ary predicate $P_f$, such that the models of $\varphi$ are in one-to-one correspondence with the models of $\psi$. Hence, for the remainder of our discussion, we will assume without loss of generality that the vocabulary is relational, i.e. has only predicate symbols. The set of all **FO** sentences over the vocabulary $\Sigma$ with equality as a special interpretted predicate (the interpretation always being the identity relation) is denoted $\mathbf{FO}_\Sigma$. When we talk about **FO** over the vocabulary $\Sigma$ without equality as a special interpretted predicate, we will explicitly refer it to it as '$\mathbf{FO}_\Sigma$ without equality'. The set of all **BSR** sentences over $\Sigma$ is denoted $\mathbf{BSR}_\Sigma$. A formula $\varphi(x_1, x_2, \ldots x_n)$ is said to be in prenex conjunctive normal form (PCNF) if it has the syntactic form $Q_1 y_1 \, Q_2 y_2 \ldots Q_k y_k \, \phi(y_1, y_2, \ldots y_k, x_1, x_2, \ldots x_n)$, where each $Q_i \in \{\forall, \exists\}$ and $\phi(y_1, y_2, \ldots y_k, x_1, x_2, \ldots x_n)$ is a conjunction of clauses, with each clause being a disjunction of literals, and each literal being either an instance of a predicate or its negation. It is well known that every **FO** formula is semantically equivalent to one in PCNF. Hence, we will focus our attention only on PCNF formulas. Since the satisfiability question for a **FO** formula can be reduced to the satisfiability of a corresponding **FO** sentence, we will focus only on **FO** sentences in the sequel.

Given a relational vocabulary $\Sigma$, a $\Sigma$-*structure* $M$ consists of a *universe (or domain)* of elements $\mathcal{U}_M$, and an *interpretation* $P^M$ of every predicate $P \in \Sigma$. The interpretation of a $k$-ary predicate over $\mathcal{U}_M$ is a subset of $(\mathcal{U}_M)^k$, giving all the tuples in $(\mathcal{U}_M)^k$ for which $P^M$ evaluates to True. Given a sentence $\varphi$ over the vocabulary $\Sigma$, a $\Sigma$-structure $M$ is said to be a *model* of $\varphi$, denoted $M \models \varphi$, if $\varphi$ evaluates to True when evaluated over $M$. If $M$ is a $\Sigma$-structure, the substructure $M'$ of $M$ generated by $\mathcal{U}_{M'} \subseteq \mathcal{U}_M$ consists of the universe $\mathcal{U}_{M'}$ and the restriction of the interpretation $P^M$ of each predicate $P$ on $\mathcal{U}_{M'}$. Given two $\Sigma$-structures $M_1$ and $M_2$ with the same universe, and a set of predicates $\sigma \subseteq \Sigma$, we say that $M_1 \mid_\sigma = M_2 \mid_\sigma$ if the interpretation of all predicates in $\sigma$ are the same in $M_1$ and $M_2$

Consider a **FO** sentence $\varphi$ in PCNF. It is well known that if the quantifier prefix of $\varphi$ consists of only $\forall$ quantifiers (such sentences are called $\forall^*$ sentences), then $\varphi$ is preserved in substructures. In other words, if $M \models \varphi$ and if $M'$ is a substructure of $M$, then $M' \models \varphi$ as well. The Loś-Tarski theorem [2] states that the converse also holds, i.e. every **FO** sentence that is preserved in substructures is semantically equivalent to a $\forall^*$ sentence. Generalizing the syntactic requirements, if we allow the quantifier prefix to be a string in $\exists^*\forall^*$, we get a sentence in the Bernays-Schönfinkel-Ramsey (**BSR**) class, also called the Effectively Propositional (**EPR**) class. Let $\mathcal{B}$ denote the length of the $\exists^*$ prefix of $\varphi$. It is not hard to see that $\varphi$ is preserved in substructures modulo a bounded "core". Specifically, if $M \models \varphi$, then there exists a "core" $\mathcal{V}_M \subseteq \mathcal{U}_M$ containing at most $\mathcal{B}$ elements such that every substructure $M'$ of $M$ that contains $\mathcal{V}_M$ is also a model of $\varphi(\mathbf{a})$. The equivalent of the Loś-Tarski theorem for **BSR** can also be obtained, and will be discussed in Section 3.

It follows from the above discussion that every $\exists^*\forall^*$ sentence has either finite or co-finite spectrum. The converse, however, is not true. Indeed, there exist **FO** sentences with co-finite spectra that neither belong to **BSR** nor are preserved under substructures modulo a bounded core. For example, consider $\psi \equiv \forall x \exists y P(x,y)$. The spectrum of $\psi$ is the entire set of natural numbers, and is hence co-finite. Yet, $\psi$ is neither in **BSR**, nor is it preserved in substructures modulo a bounded core. A natural question to ask, therefore, is whether the syntactic class **BSR** or the semantic property of "preservation in substructures modulo a bounded core" can be generalized in a natural way to describe the set of all **FO** sentences with finite or co-finite spectra in the limit. Such a generalization can be quite useful in the study and application of fragments of **FO** with finite or co-finite spectra (the widely used **BSR** class being just one of them). From a theoretical standpoint, this study can facilitate a better understanding of the semantic relation between fragments of **FO** with finite or co-finite spectra. It can also provide insights into how small models of formulae belonging to certain fragments can be constructed. Since the satisfiability problem for every fragment of **FO** with finite or co-finite spectra is decidable, identifying syntactic fragments with such spectral properties is useful in practical applications as well.

The primary contributions of this paper are as follows. We present a parameterized generalization of the "preservation in substructures modulo a bounded core" property that gives us progressively more expressive fragments of **FO** with finite or co-finite spectra, culminating in the set of all **FO** sentences with finite or co-finite spectra in the limit. We show that a large number of known decidable fragments of **FO** can be located within the fragments of this generalization. We also provide a parameterized syntactic generalization of **BSR** that gives us progressively more expressive fragments of **FO** starting from **BSR**. There has been recent work ([8]) on extending **BSR** to order-sorted logic. Our syntactic generalization is an unsorted logic and is orthogonal to the work in [8]. We study the relation between this syntactic generalisation and the parametrized semantic generalisation mentioned above. We show that the syntactic generalization enjoys special spectral properties that lends itself to interesting applications like SMT solving and bounded model checking where **BSR** has been hitherto used.

## 2 The $\mathbf{EBS}_\Sigma(\sigma)$ class

**Definition 1 (The $\mathbf{EBS}_\Sigma(\sigma)$ class)** *Let $\varphi$ be an $\mathbf{FO}_\Sigma$ sentence. Then $\varphi$ is said to have the **Extensible Bounded Sub-Model** property preserving $\sigma \subseteq \Sigma$ if the following holds: There exists a least finite cardinal $\mathcal{B}$ (called the bound for $\varphi$ - dependent on $\varphi$ and $\sigma$ in general) such that for every $\Sigma$−structure $M$, if $M \models \varphi$, then there exists a sub-structure $M_1$ of $M$ such that*

1. $|\mathcal{U}_{M_1}| \leq \mathcal{B}$

2. *For every extension $M_2$ of $M_1$ within $M$, there exists a $\Sigma$−structure $M'_2$ such that $\mathcal{U}_{M_2} = \mathcal{U}_{M'_2}$, $M'_2 \models \varphi$ and $M'_2|_\sigma = M_2|_\sigma$.*

*The set of all $\mathbf{FO}_\Sigma$ sentences having the extensible bounded sub-model property preserving $\sigma$ is called the $\mathbf{EBS}_\Sigma(\sigma)$ class.*

We will encounter examples of formulae in the above family of classes in the forthcoming sections. We organize our study of $\mathbf{EBS}_\Sigma(\sigma)$ into three parts in the sequel: (a) The case when $\sigma = \Sigma$ (b) The case when $\sigma$ is not necessarily restricted to $\Sigma$ (c) The case when $\sigma = \emptyset$.

## 3 $\mathbf{EBS}_\Sigma(\Sigma)$ as a semantic characterization of $\mathbf{BSR}_\Sigma$

**Theorem 1** *A sentence $\phi \in \mathbf{EBS}_\Sigma(\Sigma)$ iff there exists a sentence $\psi \in \mathbf{BSR}_\Sigma$ equivalent to $\phi$.*

*Proof*:
Given a $\mathbf{BSR}_\Sigma$ sentence $\psi$, from the proof of the small model property of $\mathbf{BSR}_\Sigma$ as given in [2], it can be seen that $\psi \in \mathbf{EBS}_\Sigma(\Sigma)$ with bound $\mathcal{B}$ being atmost the number of existential quantifiers of $\psi$. Then since $\mathbf{EBS}_\Sigma(\Sigma)$ is a semantic clss, if $\psi$ is equivalent to $\phi$, then $\phi$ also belongs to $\mathbf{EBS}_\Sigma(\Sigma)$.
Consider $\phi \in \mathbf{EBS}_\Sigma(\Sigma)$ in PCNF. W.l.o.g. let

$$\phi = \forall \mathbf{z}_0 \exists v_1 \forall \mathbf{z}_1 \exists v_2 \forall \mathbf{z}_2 \ldots \exists v_r \forall \mathbf{z}_r \xi(\mathbf{z}, v_1 \ldots, v_r)$$

where $\mathbf{z} = (\mathbf{z}_0, \ldots, \mathbf{z}_r)$ and $\xi$ is quantifier-free. Let $S_\mathbf{z}$ be the set of variables of $\mathbf{z}$. Introduce fresh variables $x_1, \ldots, x_\mathcal{B}$ where $\mathcal{B}$ is the bound for $\phi$ and let $\mathcal{V} = \{x_1, \ldots, x_\mathcal{B}\} \cup S_\mathbf{z}$. Consider $\psi$ as below.

$\psi = \exists x_1 \ldots \exists x_\mathcal{B} \forall \mathbf{z}_0 \forall \mathbf{z}_1 \ldots \forall \mathbf{z}_r (\chi)$ where
$\chi = (\bigvee_{u_1 \in \mathcal{V}} \cdots \bigvee_{u_r \in \mathcal{V}} \xi(\mathbf{z}, v_1 \mapsto u_1, \ldots, v_r \mapsto u_r))$
Here $\xi(\mathbf{z}, v_1 \mapsto u_1, \ldots, v_r \mapsto u_r)$ is the formula obtained by replacing every occurence of $v_i$ in $\xi$ by $u_i$.
If $M \models \phi$ and $|\mathcal{U}_M| \leq \mathcal{B}$, then assign values to $x_1 \ldots x_\mathcal{B}$ such that each $a \in \mathcal{U}_M$ is assigned to some $x_i$. Consider an instantiation $\mathcal{Z}$ of $\mathbf{z}$ from $\mathcal{U}_M$. Since $M \models \phi$, there exist values $d_1, \ldots, d_r$ for $v_1, \ldots, v_r$ such that $M \models \xi(\mathcal{Z}, d_1, \ldots, d_r)$. Each $d_i$ is the value assigned to some $u_i \in \{x_1, \ldots, x_\mathcal{B}\}$. Then $M \models \xi(\mathbf{z}, v_1 \mapsto u_1, \ldots, v_r \mapsto u_r)$ and so $M \models \chi$. Since $\mathcal{Z}$ was arbitrary, $M \models \psi$.
If $M \models \phi$ and $|\mathcal{U}_M| \geq \mathcal{B}$, then there exists $M_1 \subseteq M$ satisfying the $\mathbf{EBS}_\Sigma(\Sigma)$ conditions. Assign values to $x_1 \ldots x_\mathcal{B}$ s.t. each $x_i$ is assigned some $a \in \mathcal{U}_{M_1}$ and each $a \in \mathcal{U}_{M_1}$ is assigned to some $x_j$. Consider an instantiation $\mathcal{Z}$ of $\mathbf{z}$. Let the set of elements of $\mathcal{Z}$ be $\mathcal{T}$. Then consider the substructure $M_2$ of $M$ generated by $\mathcal{U}_{M_1} \cup \mathcal{T}$. Since $M_1 \subseteq M_2 \subseteq M$, $M_2 \models \phi$. Then there exist values $d_1, \ldots, d_r$ for $v_1, \ldots, v_r$ such that $M_2 \models \xi(\mathcal{Z}, d_1, \ldots, d_r)$. Observe that since $M_2 \subseteq M$, $M \models \xi(\mathcal{Z}, d_1, \ldots, d_r)$. Now since $d_i \in \mathcal{U}_{M_1} \cup \mathcal{T}$, it is the value assigned to some $u_i \in \mathcal{V}$. Then since $M \models \xi(\mathcal{Z}, d_1, \ldots, d_r)$, $M \models \xi(\mathcal{Z}, v_1 \mapsto u_1, \ldots, v_r \mapsto u_r)$ and so $M \models \chi$. Since $\mathcal{Z}$ was arbitrary, $M \models \psi$.

Now suppose $M \models \psi$. Then there exists a set of values $S \subseteq \mathcal{U}_M$ such that the $x_i$'s in $\psi$ get their values from this set and for all instantiations from $\mathcal{U}_M$ of $\mathbf{z}$, $M \models \chi$. Then consider the instantiation $\mathcal{Z}$ of $\mathbf{z}$. Since $M \models \chi$ for $\mathcal{Z}$, there exists some disjunct $\xi(\mathcal{Z}, v_1 \mapsto u_1; \ldots; v_r \mapsto u_r)$ which is True in $M$. Now $u_i \in \mathcal{V}$ and hence is either some $x_j$ or a universal variable $z$. Then if $u_i$ is $x_j$, let $d_i$ be the value (from $S$) assigned to $x_j$. If $u_i$ is the variable $z$, let $d_i$ be the value assigned to $z$ in $\mathcal{Z}$. Then $M \models \xi(\mathcal{Z}, d_1 \ldots, d_r)$. Then for the instantiation $\mathcal{Z}$, choosing $v_i = d_i$, we see that the matrix of $\phi$ becomes True in $M$. Doing likewise for all instantiations of $\mathbf{z}$, we conclude that $M \models \phi$.
From the above then, $M \models \phi \leftrightarrow M \models \psi$ i.e. $\phi$ is equivalent to $\psi$, the latter being a sentence in $\mathbf{BSR}_\Sigma$. ∎

**Some consequences of Theorem 1**

1. *Loś-Tarski Theorem as a special case:* Consider the class of $\mathbf{FO}_\Sigma$ sentences $\phi$ which are preserved under substructures. Then we can see that $\phi \in \mathbf{EBS}_\Sigma(\Sigma)$ with $\mathcal{B} = 0$. Then from the proof of the Theorem 1, we can see that $\phi$ is equivalent to a $\mathbf{BSR}_\Sigma$ sentence $\psi$ where the number of existential quantifiers is 0. In other words $\phi$ is equivalent to a $\forall^*$ sentence. The converse, namely, that a $\forall^*$ sentence is closed under substructures is well-known. We thus get the Loś-Tarski Theorem as a special case of Theorem 1. The proof of Theorem 1 serves as an alternate proof of the Loś-Tarski Theorem.

   The $\mathbf{EBS}_\Sigma(\Sigma)$ property can be viewed as preservation under substructures *modulo a bounded 'core'* i.e. all substructures of a model which contain the core of the model are also models. The $\exists^*$ quantifiers in the equivalent $\mathbf{BSR}_\Sigma$ formula of an $\mathbf{EBS}_\Sigma(\Sigma)$ formula can be seen to assert the existence of a bounded extensible core.

2. $\mathbf{BSR}_\Sigma$ *semantically subsumes other classes:* Theorem 1 helps to show right away that some classes of $\mathbf{FO}_\Sigma$ formulae are semantically contained in $\mathbf{BSR}_\Sigma$. The Löwenheim class is the class of all $\mathbf{FO}_\Sigma$ sentences over $\Sigma$, without equality where $\Sigma$ contains only unary predicates. The Löwenheim class with equality - denoted $\mathbf{L}_\Sigma$ - is the class of all $\mathbf{FO}_\Sigma$ sentences over $\Sigma$, possibly containing equality where $\Sigma$ contains only unary predicates (Whenever we will talk about just the Löwnheim class, we will refer to it as '$\mathbf{L}_\Sigma$ without equality'). By an Ehrenfeucht-Fraïssé game argument (see [2], pp. 259), one can show that if $\phi \in \mathbf{L}_\Sigma$, then $\phi \in \mathbf{EBS}_\Sigma(\Sigma)$ with the bound $\mathcal{B} = q \cdot 2^m$ where $q$ is the length of the quantifier prefix of $\phi$ and $m$ is the number of unary predicates ($m = |\Sigma|$). By theorem 1, we immediately know that for $\phi$, there is an equivalent $\mathbf{BSR}_\Sigma$ sentence $\psi$ whose number of existential quantifiers is $\mathcal{B}$. Since $\Sigma$ contains only unary predicates, this shows us that $\mathbf{L}_\Sigma$ is semantically the same as the $\text{BSR}_\Sigma$ class, the latter being a proper syntactic subset of the former.

Other known classes which can be seen to be subsumed semantically by $\text{BSR}_\Sigma$ include $\text{FO}_\Sigma[1]$ and $\text{FO}_\Sigma[1]$ with counting which are respectively the class of $\text{FO}_\Sigma$ sentences having only one variable and the class of $\text{FO}_\Sigma$ sentences with counting having only one variable. These are contained inside $\text{L}_\Sigma$ and hence are semantically within $\text{BSR}_\Sigma$.

3. *Translation from $\text{EBS}_\Sigma(\Sigma)$ to $\text{BSR}_\Sigma$:* The proof not only shows us that for a sentence $\phi \in \text{EBS}_\Sigma(\Sigma)$, there exists an equivalent $\text{BSR}_\Sigma$ sentence, but also gives us the sentence ($\psi$ in the proof). It gives a uniform translation scheme from the whole of $\text{EBS}_\Sigma(\Sigma)$ to $\text{BSR}_\Sigma$. As for the size of $\psi$, the number of $\exists$ variables is $\mathcal{B}$, the number of $\forall$ variables is atmost $k + 1$ where $k$ is the number of $\forall$ variables in $\phi$ and the size of the matrix is $O((\mathcal{B} + k + 1)^r \cdot |\xi|)$ where $|\xi|$ is the length of $\xi$ and where $r$ (number of existential quantifiers) and $\xi$ are as in the proof.

4. *Optimal translation:* The proof also shows us that for a sentence $\phi \in \text{EBS}_\Sigma(\Sigma)$ whose bound is $\mathcal{B}$, there cannot exist an equivalent $\text{BSR}_\Sigma$ sentence with $\mathcal{B}'$ $\exists$ quantifiers where $\mathcal{B}' < \mathcal{B}$. Otherwise, we would violate the minimality of $\mathcal{B}$.

5. *Finding the bound $\mathcal{B}$:* $\text{EBS}_\Sigma(\Sigma)$ is exactly the class of all those sentences which are equivalent to formulae in $\text{BSR}_\Sigma$. This syntactic characterization of $\text{EBS}_\Sigma(\Sigma)$ as $\text{BSR}_\Sigma$, the latter being a syntactic class for which membership is decidable, shows us that the membership problem for $\text{EBS}_\Sigma(\Sigma)$ is recursively enumerable (We show later that the membership problem for $\text{EBS}_\Sigma(\Sigma)$ is undecidable though in general). Suppose we knew that $\phi$ belongs to $\text{EBS}_\Sigma(\Sigma)$, but did not know the bound $\mathcal{B}$ for $\phi$. A naive approach would be to try to prove equivalence of $\phi$ with each $\text{BSR}_\Sigma$ formula with $k$ $\exists$ quantifiers, for increasing values of $k$. The above proof however shows that for each $k$, it is sufficient to construct *exactly one* $\text{BSR}_\Sigma$ formula with which the equivalence of $\phi$ needs to be checked. Following is the procedure of finding $\mathcal{B}$ in detail:

For each $\mathcal{B} \geq 0$, we construct the formula $\psi$ as shown in the proof. Call this formula $\psi_\mathcal{B}$. We then construct the formula $\Xi_\mathcal{B} = (\phi \leftrightarrow \psi_\mathcal{B})$. We then interleave the procedures which check for the validity of $\Xi_\mathcal{B}$'s for $\mathcal{B} \geq 0$. If $\phi$ indeed belongs to $\text{EBS}_\Sigma(\Sigma)$, then one of these procedures is guaranteed to terminate. The least $\mathcal{B}$ for which $\Xi_\mathcal{B}$ is valid gives the bound $\mathcal{B}$ for $\phi$ (since, if the bound for $\phi$ was lower, say $\mathcal{B}'$, then from the above proof, we know that $\Xi_{\mathcal{B}'}$ would have been valid thus violating the minimality of $\mathcal{B}$). It also gives an equivalent formula for $\phi$, namely $\psi_\mathcal{B}$, with the least number of $\exists^*$ quantifiers. This procedure thus requires us to construct *exactly one* $\text{BSR}_\Sigma$ formula having $\mathcal{B}$ $\exists$ quantifiers for each $\mathcal{B}$ in order to find an equivalent $\text{BSR}_\Sigma$ formula for $\phi$. It just means that $\text{BSR}_\Sigma$ formulae other than the $\psi_\mathcal{B}$s need not be enumerated at all.

6. *Uses in SAT and model checking:* An advantage of converting an $\text{EBS}_\Sigma(\Sigma)$ formula $\phi$ to its equivalent $\text{BSR}_\Sigma$ formula $\psi$ is that $\phi$ is satisfiable iff $\psi$ is satisfiable. There has been a lot of recent work [3] on decision procedures for the $\text{EPR}$ class - which is the same as $\text{BSR}$. While the fact that $\phi \in \text{EBS}_\Sigma(\Sigma)$ shows that the SAT-problem for $\text{EBS}_\Sigma(\Sigma)$ is decidable and the bound $\mathcal{B}$ for $\phi$ gives a way of checking the satisfiability of $\phi$, we can leverage recent advances in decision procedures for $\text{BSR}$ to check for satisfiability of $\phi$.

Likewise the model checking problem for $\phi$ (given a $\Sigma$−structure $M$, is it true that $M \models \phi$) would be the same as that for $\psi$. Hence the power of model checkers optimized for $\text{BSR}$ could be leveraged.

## 4  Properties of the $\text{EBS}_\Sigma(\sigma)$ classes

The $\text{EBS}_\Sigma(\Sigma)$ property required that any model $M$ of a formula $\phi \in \text{EBS}_\Sigma(\Sigma)$ is such that $\phi$ is also True in any substructure of $M$ which contains the 'bounded core' of $M$. While $\text{EBS}_\Sigma(\Sigma)$ required the preservation of the interpretation of *all* the predicates of $\Sigma$, $\text{EBS}_\Sigma(\sigma)$ relaxes this requirement. It insists that the interpretation of only the predicates of a specified subset $\sigma$ of $\Sigma$ be preserved. The $\Sigma \setminus \sigma$ predicates are free to be re-interpreted in any way by which $\phi$ becomes True in the resulting structure. Specifically, condition 2 in Definition 1 in section 2 does not require the extension $M_2$ to satisfy $\varphi$. Instead, it requires the existence of a model $M_2'$ which possibly differs from $M_2$ in the interpretation of predicates in $\Sigma \setminus \sigma$.

As an example, consider $\varphi = \forall x \exists y P(x, y)$. Here $\Sigma = \{P\}$. If we consider $\sigma = \emptyset$, then by choosing $\mathcal{B} = 1$, for any model $M$ of size $\geq \mathcal{B}$, take $M_1$ to be the substructure generated by any one element of $M$. Then for any extension $M_2$ of $M_1$ within $M$, construct $M_2'$ by re-interpreting $P$ to be the identity relation. Clearly then $M_2' \models \varphi$. Thus $\varphi \in \text{EBS}_\Sigma(\emptyset)$.

If however $\sigma = \Sigma$, then consider the model $M$ in which $\mathcal{U}_M = \mathbb{N}$ and $P(i, j) = $ True iff $j = i + 1$. Suppose $\varphi \in \text{EBS}_\Sigma(\Sigma)$. Then there exists a sub-structure $M_1$ of $M$ of some bounded size say $\mathcal{B}$ such that $M_1 \models \varphi$. Consider the highest number $i^*$ in $\mathcal{U}_{M_1}$. Then there exists $j^* \in \mathcal{U}_{M_1}$ such that $P(i^*, j^*)$ is True in $M_1$. But since $M_1 \subseteq M$, $j^* = i^* + 1$. This contradicts the assumption that $i^*$ is the highest number in $\mathcal{U}_{M_1}$. This shows that $\varphi \notin \mathbf{EBS}_\Sigma(\Sigma)$.

Let us define the family $\mathcal{EBS}_\Sigma = \{\mathbf{EBS}_\Sigma(\sigma) | \sigma \subseteq \Sigma\}$. In the section below, we look into set inclusions between the different classes in $\mathcal{EBS}_\Sigma$, and show that the poset $(\mathcal{EBS}_\Sigma, \subseteq)$ has a lattice structure to it.

## 4.1 Lattice Structure of $(\mathcal{EBS}_\Sigma, \subseteq)$

**Theorem 2** *1. If $\sigma_1 \subseteq \sigma_2 \subseteq \Sigma$, then $\mathbf{EBS}_\Sigma(\sigma_2) \subseteq \mathbf{EBS}_\Sigma(\sigma_1)$.*

*2. If $\Sigma$ contains an arity $\geq 2$ predicate, then given $\sigma_1, \sigma_2 \in \Sigma$, if $\sigma_1 \setminus \sigma_2$ is non-empty, then $\mathbf{EBS}_\Sigma(\sigma_2) \setminus \mathbf{EBS}_\Sigma(\sigma_1)$ is also non-empty.*

*3. If $\Sigma$ contains only unary predicates, then $\mathbf{EBS}_\Sigma(\sigma_1) = \mathbf{EBS}_\Sigma(\sigma_2)$ for any $\sigma_1, \sigma_2 \subseteq \Sigma$.*

*Proof:*
Statement (1) above is easy to see. Consider $\phi \in \mathbf{EBS}_\Sigma(\sigma_2)$. Then $\phi$ satisfies condition 2 of Definition 1 for $\sigma = \sigma_2$. Since $\sigma_1 \subseteq \sigma_2$, $M'_2|_{\sigma_2} = M_2|_{\sigma_2}$ implies that $M'_2|_{\sigma_1} = M_2|_{\sigma_1}$. Then $\phi$ satisfies the condition 2 for $\sigma = \sigma_1$ as well. The first condition of Definition 1 is independent of $\sigma$. Then $\phi \in \mathbf{EBS}_\Sigma(\sigma_1)$. Then $\mathbf{EBS}_\Sigma(\sigma_2) \subseteq \mathbf{EBS}_\Sigma(\sigma_1)$.
We now prove statement (2) above. We consider the following cases.
1) $\sigma_1 \setminus \sigma_2$ contains an arity $\geq 2$ predicate. Let $P$ be such a predicate of arity $k \geq 2$.
We use the same idea as in the example above to construct a formula $\phi \in \mathbf{EBS}_\Sigma(\sigma_2) \setminus \mathbf{EBS}_\Sigma(\sigma_1)$. Construct $\phi$ as follows.

$$\phi = (\bigwedge_{Q \in \Sigma \setminus \{P\}} \forall \mathbf{z}_Q Q(\mathbf{z}_Q)) \wedge (\forall x \exists y P(x, y, y, \ldots, y))$$

where $\mathbf{z}_Q$ is a vector of variables where the length of the vector is equal to the arity of $Q$. Call the first and second conjuncts in the above formula as $\alpha$ and $\beta$ respectively. We first show that $\phi \in \mathbf{EBS}_\Sigma(\sigma_2)$. Consider a model $M$ of $\phi$ of size $\geq 1$. Consider any element $a \in \mathcal{U}_M$ and let $M_1$ be the substructure of $M$ generated by $\{a\}$. Consider any $M_2$ such that $M_1 \subseteq M_2 \subseteq M$. Consider $M'_2$ obtained from $M_2$ in which the interpretation of all the $\Sigma \setminus \{P\}$ predicates are retained as in $M_2$ (this will mean that the interpretation of the predicates of $\sigma_2$ will also be preserved). But for $P$, since $P \notin \sigma_2$, re-interpret $P$ to be 'fully' True, i.e. in $M'_2$, $P$ is True for all values of its arguments. Now since $\alpha$ is a $\forall^*$ formula over the vocabulary $\Sigma \setminus \{P\}$, it is True in all sub-structures of $M|_{\Sigma \setminus \{P\}}$. Then since $M'_2|_{\Sigma \setminus \{P\}} \subseteq M|_{\Sigma \setminus \{P\}}$, $M'_2|_{\Sigma \setminus \{P\}} \models \alpha$ and hence $M'_2 \models \alpha$. Further since $P$ is fully True in $M'_2$, $M'_2 \models \beta$. Thus $M'_2 \models \phi$. This shows $\phi \in \mathbf{EBS}_\Sigma(\sigma_2)$ with $\mathcal{B} = 1$.

We now show that $\phi \notin \mathbf{EBS}_\Sigma(\sigma_1)$. Consider the model $M$ in which $\mathcal{U}_M = \mathbb{N}$. Let all the predicates in $\Sigma \setminus \{P\}$ be fully True in $M$. Let $P(x, y_1, \ldots, y_{k-1}) = \text{True iff } y_1 = y_2 = \cdots = y_{k-1} = y$ and $y = x + 1$. Consider any sub-structure $M_1$ of $M$ of size $\leq \mathcal{B}$ for some $\mathcal{B}$. Let $M_2 = M_1$ and consider any sub-structure $M'_2$ obtained from $M_2$ by possibly re-interpreting the predicates of $\Sigma \setminus \sigma_1$ but keeping the interpretations of the $\sigma_1$ predicates the same. If $M'_2 \models \phi$, then $M'_2 \models \beta$. Let $i^*$ be the highest number in $\mathcal{U}_{M'_2} = \mathcal{U}_{M_2}$. Since $M'_2 \models \beta$, there exists $j^* \in \mathcal{U}_{M_2}$ such that $M'_2 \models P(i^*, j^*, \ldots, j^*)$. But since $P \in \sigma_1$, $M'_2|_{\{P\}} = M|_{\{P\}}$ and hence $j^* = i^* + 1$. But that contradicts the assumption that $i^*$ is the highest number in $\mathcal{U}_{M_2}$. This shows that $\phi \notin \mathbf{EBS}_\Sigma(\sigma_1)$.

2) $\sigma_1 \setminus \sigma_2$ contains only unary predicates. Then $P \notin \sigma_1 \setminus \sigma_2$. Let $U \in \sigma_1 \setminus \sigma_2$. In order to save space, below $\mathbf{P}(x, y)$ is a shorthand for $P(x, y, \ldots, y)$.
Construct $\phi$ as follows.

$$\begin{aligned}
\alpha_1 &= \forall x \forall y_1 \ldots \forall y_{k-1}(P(x, y_1, \ldots, y_{k-1}) \to \\
&\quad ((y_1 = y_2) \wedge \ldots \wedge (y_{k-2} = y_{k-1}))) \\
\alpha_2 &= \forall x \forall y ((x = y) \vee (\mathbf{P}(x, y) \oplus \mathbf{P}(y, x))) \wedge \\
&\quad \forall x \forall y \forall z (\mathbf{P}(x, y) \wedge \mathbf{P}(y, z)) \to \mathbf{P}(x, z) \\
\beta &= (\bigwedge_{Q \in \Sigma \setminus \{P, U\}} \forall \mathbf{z}_Q Q(\mathbf{z}_Q)) \\
\text{succ}(x, y) &= (x \neq y) \wedge \mathbf{P}(x, y) \wedge \\
&\quad \forall z ((z \neq x \wedge z \neq y \wedge \mathbf{P}(x, z)) \to \mathbf{P}(y, z)) \\
\gamma &= \forall x \forall y (\text{succ}(x, y) \to (U(x) \leftrightarrow \neg U(y))) \\
\phi &= \alpha_1 \wedge \alpha_2 \wedge \beta \wedge \gamma
\end{aligned}$$

Above $\oplus$ means XOR. The condition $\alpha_1 \wedge \alpha_2$ states that $P$ is essentially a linear order. $\gamma$ states that the elements of the linear order are alternately labelled with $U$ and $\neg U$.
We first show that $\phi \in \mathbf{EBS}_\Sigma(\sigma_2)$. Consider any model $M$ of $\phi$ of size $\geq 1$. Let $a \in \mathcal{U}_M$. Then consider the substructure $M_1$ generated by $\{a\}$. Consider $M_2$ such that $M_1 \subseteq M_2 \subseteq M$. Consider $M'_2$ obtained from $M_2$ by interpreting all the predicates of $\Sigma \setminus \{U\}$ as in $M_2$ but in which $U$ is interpreted such that the elements of the linear order defined by $P$ in $M'_2$ are alternately labelled with $U$ and $\neg U$ (the linear order amongst the elements of $\mathcal{U}_{M'_2}$ is the same in $M'_2$ as the linear order amongst them in $M$). Clearly then $M'_2 \models \gamma$. Now since $\alpha_1 \wedge \alpha_2 \wedge \beta$ is a $\forall^*$ sentence over $\Sigma \setminus \{U\}$ and since $M'_2|_{\Sigma \setminus \{U\}} \subseteq M|_{\Sigma \setminus \{U\}}$, we have $M'_2|_{\Sigma \setminus \{U\}} \models (\alpha_1 \wedge \alpha_2 \wedge \beta)$ and hence $M'_2 \models (\alpha_1 \wedge \alpha_2 \wedge \beta)$. Thus $M'_2 \models \phi$. This shows that $\phi \in \mathbf{EBS}_\Sigma(\sigma_2)$ with $\mathcal{B} = 1$.
We now show that $\phi \notin \mathbf{EBS}_\Sigma(\sigma_1)$. Consider a model $M$ of $\phi$ in which $\mathcal{U}_M = \mathbb{N}$, all predicates in $\Sigma \setminus \{P, U\}$ are fully True, $P$ is essentially the usual '<' linear order and the elements of $\mathcal{U}_M$ are alternately labelled with $U$ and $\neg U$. Now consider any sub-structure $M_1$ of $M$ of size $\leq \mathcal{B}$ for some $\mathcal{B}$. Let the number of '$\neg U$' elements in $\mathcal{U}_{M_1}$ be $m$. Then consider any subset $\mathcal{U}^* \subseteq \mathcal{U}_M$ such that (i) $\mathcal{U}_{M_1} \subsetneq \mathcal{U}^*$ and $\mathcal{U}^* \setminus \mathcal{U}_{M_1}$ contains only those elements from $\mathcal{U}_M$ which are labelled with a 'U' in $M$ and (ii) if the number of 'U' elements of $M$ in $\mathcal{U}^*$ is $n$, then $n - m \geq 2$. Let $M_2$ be the substructure of $M$ generated by $\mathcal{U}^*$. Then $M_1 \subseteq M_2 \subseteq M$. Now consider any sub-structure $M_2'$ obtained from $M_2$ by possibly re-interpreting the predicates of $\Sigma \setminus \sigma_1$ but keeping the interpretations of the $\sigma_1$ predicates the same. If $M_2' \models \phi$, then $M_2' \models (\alpha_1 \wedge \alpha_2 \wedge \gamma)$. Then $P$ would impose a linear order on the elements of $\mathcal{U}_{M_2'}$ (This linear order would be the same as that imposed on $\mathcal{U}_{M_2'}$ by the interpretation of $P$ in $M$ if $P \in \sigma_1$ and could be different from that imposed on $\mathcal{U}_{M_2'}$ by the interpretation of $P$ in $M$ if $P \notin \sigma_1$. But in either case, it would be a linear order). Now since $U \in \sigma_1$, the interpretation of $U$ is the same in $M$ and $M_2'$ and hence the number of elements labelled by $U$ and $\neg U$ in $M_2'$ are $n$ and $m$ respectively. Since $M_2' \models \gamma$, the elements of $\mathcal{U}_{M_2'}$ are alternately labelled by $U$ and $\neg U$ so that $|n - m| \leq 1$. But this contradicts the fact that $n - m \geq 2$. This shows that $\phi \notin \mathbf{EBS}_\Sigma(\sigma_1)$.

Thus in all cases we have shown that $\mathbf{EBS}_\Sigma(\sigma_2) \setminus \mathbf{EBS}_\Sigma(\sigma_1)$ is non-empty.

The proof of statement 3 of theorem 2 is easy to see. If $\Sigma$ contains only unary predicates, then $\mathbf{FO}_\Sigma = \mathbf{L}_\Sigma$. Then as seen from the consequences of Theorem 1 in section 3, $\mathbf{FO}_\Sigma \subseteq \mathbf{EBS}_\Sigma(\Sigma)$. Then for any $\sigma \subseteq \Sigma$, since $\mathbf{EBS}_\Sigma(\sigma) \subseteq \mathbf{FO}_\Sigma$, we have $\mathbf{EBS}_\Sigma(\sigma) \subseteq \mathbf{EBS}_\Sigma(\Sigma)$. But from statement 1 of theorem 2, we have $\mathbf{EBS}_\Sigma(\Sigma) \subseteq \mathbf{EBS}_\Sigma(\sigma)$. Thus $\mathbf{EBS}_\Sigma(\sigma) = \mathbf{EBS}_\Sigma(\Sigma)$ for any $\sigma \subseteq \Sigma$. ∎

**Corollary 1** *If $\Sigma$ contains a predicate of arity $\geq 2$, then $(\mathcal{EBS}_\Sigma, \subseteq)$ is a complete distributive lattice which is isomorphic to the powerset lattice $(\wp(\Sigma), \subseteq)$. If $\Sigma$ contains only unary predicates, then $(\mathcal{EBS}_\Sigma, \subseteq)$ collapses to the singleton lattice.*

*Proof:*
Suppose $\Sigma$ contains an arity $\geq 2$ predicate. Consider $(\mathcal{EBS}_\Sigma, \subseteq)$. Now if $\sigma_1 \neq \sigma_2$, then from statement 2 of Theorem 2, $\mathbf{EBS}_\Sigma(\sigma_1) \neq \mathbf{EBS}_\Sigma(\sigma_2)$. Then we can define the function $f : \mathcal{EBS}_\Sigma \to \wp(\Sigma)$ given by $f(\mathbf{EBS}_\Sigma(\sigma)) = \Sigma \setminus \sigma$. If $f(\mathbf{EBS}_\Sigma(\sigma_1)) = f(\mathbf{EBS}_\Sigma(\sigma_2))$ then $\Sigma \setminus \sigma_1 = \Sigma \setminus \sigma_2$ or $\sigma_1 = \sigma_2$ and hence $\mathbf{EBS}_\Sigma(\sigma_1) = \mathbf{EBS}_\Sigma(\sigma_2)$. Thus $f$ is injective. For every $\sigma \in \wp(\Sigma)$, $f(\mathbf{EBS}_\Sigma(\Sigma \setminus \sigma)) = \sigma$. Thus $f$ is surjective and hence bijective.

Suppose $\mathbf{EBS}_\Sigma(\sigma_1) \subseteq \mathbf{EBS}_\Sigma(\sigma_2)$. Then if $\sigma_2 \setminus \sigma_1 \neq \{\}$, then from statement 2 of Theorem 2, $\mathbf{EBS}_\Sigma(\sigma_1) \setminus \mathbf{EBS}_\Sigma(\sigma_2) \neq \{\}$ violating the assumption just made. Then $\sigma_2 \subseteq \sigma_1$ or $\Sigma \setminus \sigma_1 \subseteq \Sigma \setminus \sigma_2$ or $f(\mathbf{EBS}_\Sigma(\sigma_1)) \subseteq f(\mathbf{EBS}_\Sigma(\sigma_2))$.

Suppose $f(\mathbf{EBS}_\Sigma(\sigma_1)) \subseteq f(\mathbf{EBS}_\Sigma(\sigma_2))$. Then $\Sigma \setminus \sigma_1 \subseteq \Sigma \setminus \sigma_2$ or $\sigma_2 \subseteq \sigma_1$. Then from statement 1 of Theorem 2, $\mathbf{EBS}_\Sigma(\sigma_1) \subseteq \mathbf{EBS}_\Sigma(\sigma_2)$.

Thus $\mathbf{EBS}_\Sigma(\sigma_1) \subseteq \mathbf{EBS}_\Sigma(\sigma_2)$ iff $f(\mathbf{EBS}_\Sigma(\sigma_1)) \subseteq f(\mathbf{EBS}_\Sigma(\sigma_2))$. Thus $f$ is an isomorphism between $(\mathcal{EBS}_\Sigma, \subseteq)$ and the powerset lattice $(\wp(\Sigma), \subseteq)$.

Suppose $\Sigma$ contains only unary predicates. Then from Theorem 2 (3), we see that $\mathcal{EBS}_\Sigma$ contains only one element, namely $\mathbf{EBS}_\Sigma(\Sigma)$. ∎

The *lub*($\sqcup$) and *glb*($\sqcap$) operators in $(\mathcal{EBS}_\Sigma, \subseteq)$ are defined as: $\mathbf{EBS}_\Sigma(\sigma_1) \sqcup \mathbf{EBS}_\Sigma(\sigma_2) = \mathbf{EBS}_\Sigma(\sigma_1 \cap \sigma_2)$ and $\mathbf{EBS}_\Sigma(\sigma_1) \sqcap \mathbf{EBS}_\Sigma(\sigma_2) = \mathbf{EBS}_\Sigma(\sigma_1 \cup \sigma_2)$. Note that $\mathbf{EBS}_\Sigma(\sigma_1 \cup \sigma_2) \subseteq (\mathbf{EBS}_\Sigma(\sigma_1) \cap \mathbf{EBS}_\Sigma(\sigma_2))$ and $(\mathbf{EBS}_\Sigma(\sigma_1) \cup \mathbf{EBS}_\Sigma(\sigma_2)) \subseteq \mathbf{EBS}_\Sigma(\sigma_1 \cap \sigma_2)$.

### 4.2 $\mathbf{EBS}_\Sigma(\sigma)$ and other semantic classes

Let $\mathbf{FMP}_\Sigma$ be the set of all $\mathbf{FO}_\Sigma$ sentences $\phi$ such that if $\phi$ has a model, it also has a finite model. Let $\mathbf{FSMP}_\Sigma$ be the set of all $\mathbf{FO}_\Sigma$ sentences $\phi$ such that for every model of $\phi$, there is a finite sub-model of it. One can then show the following relations of these with $\mathbf{EBS}_\Sigma(\sigma)$.

**Lemma 1** *1. If $\Sigma$ contains only unary predicates, then $\mathbf{FO}_\Sigma = \mathbf{FMP}_\Sigma = \mathbf{FSMP}_\Sigma = \mathbf{EBS}_\Sigma(\sigma)$ for all $\sigma \subseteq \Sigma$.*

*2. If $\Sigma$ contains an arity $\geq 2$ predicate, then $\forall \, \sigma \subseteq \Sigma$,*

   *(a) $\mathbf{EBS}_\Sigma(\sigma) \subsetneq \mathbf{FMP}_\Sigma$*

   *(b) (i) $\mathbf{EBS}_\Sigma(\Sigma) \subsetneq \mathbf{FSMP}_\Sigma$ (ii) $\mathbf{EBS}_\Sigma(\sigma) \setminus \mathbf{FSMP}_\Sigma$ and $\mathbf{FSMP}_\Sigma \setminus \mathbf{EBS}_\Sigma(\sigma)$ are both non-empty for $\sigma \neq \Sigma$.*

*Proof*:
Consider Statement 1 above. If $\Sigma$ contains only unary predicates, then $\mathbf{FO}_\Sigma = \mathbf{L}_\Sigma$ (the Löwenheim class with equality). Then as seen in section 3, $\mathbf{FO}_\Sigma \subseteq \mathbf{EBS}_\Sigma(\Sigma)$. From Theorem 2(3), for all $\sigma \subseteq \Sigma$, $\mathbf{EBS}_\Sigma(\Sigma) = \mathbf{EBS}_\Sigma(\sigma) \subseteq \mathbf{FO}_\Sigma$. Then every $\mathbf{FO}_\Sigma$ formula has the finite model, finite sub-model and the $\mathbf{EBS}_\Sigma(\sigma)$ properties for each $\sigma \subseteq \Sigma$.

Consider the case when $\Sigma$ contains a predicate $P$ of arity $\geq 2$. Every $\mathbf{EBS}_\Sigma(\sigma)$ sentence is either unsatisfiable or has a finite model (of size $\mathcal{B}$ where $\mathcal{B}$ is the bound for $\phi$) and hence has the finite model property. Thus for all $\sigma \subseteq \Sigma$, $\mathbf{EBS}_\Sigma(\sigma) \subseteq \mathbf{FMP}_\Sigma$. But the converse is not true. Consider the sentence $\phi$ as below.

$$\phi = \quad \alpha_1 \wedge \alpha_2 \wedge$$
$$\forall x \forall y (\text{succ}(x, y) \to (\mathbf{P}(x, x) \leftrightarrow \neg \mathbf{P}(y, y)) \wedge$$
$$\exists x \exists y (\text{first}(x) \wedge \mathbf{P}(x, x) \wedge \text{last}(y) \wedge \neg \mathbf{P}(y, y))$$

where $\alpha_1, \alpha_2$ and succ$(x, y)$ are as in the proof of case (2) of statement 2 of Theorem 2. first$(x)$ and last$(x)$ denote that $x$ is respectively the first and last element of the linear order. Any finite model of $\phi$ can be seen to be 'essentially' a linear order with its elements alternately coloured and not coloured (an element $a$ is coloured iff $\mathbf{P}(a, a)$ is True), where the first element is coloured and the last is not. Then $\phi$ has finite models of only even cardinality. Thus $\phi$ has the finite model property. Its easy to see that $\phi$ also has the finite sub-model property. But from Lemma 2 (in the forthcoming section 5), $\phi$ cannot be in $\mathbf{EBS}_\Sigma(\emptyset)$ and hence from Theorem 2(1), it cannot be in $\mathbf{EBS}_\Sigma(\sigma)$ for any $\sigma \subseteq \Sigma$. This shows statement 2a and the second part of statement 2b(ii).

Consider statement 2b. Now by definition, all sentences in $\mathbf{EBS}_\Sigma(\Sigma)$ are also in $\mathbf{FSMP}_\Sigma$. Then considering $\phi$ above, we get statement 2b(i).

Consider $\mathbf{EBS}_\Sigma(\sigma)$. If $\Sigma \setminus \sigma$ contains a predicate $P$ of arity $\geq 2$, then consider the sentence $\phi$ as given in the proof of part (1) of statement 2 of Theorem 2. This sentence cannot have the finite submodel property. If $\Sigma \setminus \sigma$ contains only unary predicates, then consider one such unary predicate $U$. Then consider the sentence $\phi = \phi_1 \wedge \phi_2$ where $\phi_1$ asserts that $P$ is essentially a linear order and $\phi_2$ asserts that either there is no last element in the model or if there is one, then it must be labelled with $U$. Then consider an infinite linear order in which no element is labelled with $U$. One can see that while this models $\phi$, no finite sub-model of this can model $\phi$. But since $U$ lies outside $\sigma$, for any finite sub-model, one can change the interpretation of $U$ and label the last element of the finite sub-model with $U$ to make $\phi$ true in the resulting model. Hence $\phi \in \mathbf{EBS}_\Sigma(\sigma)$. This finally shows the first part of 2b(ii). ■

## 5 The $\mathbf{EBS}_\Sigma(\emptyset)$ class

In this section, we study the $\mathbf{EBS}_\Sigma(\emptyset)$ class. We firstly observe that this class subsumes some well-known syntactic classes.

### 5.1 The $\mathbf{EBS}_\Sigma(\emptyset)$ class subsumes well-known syntactic classes

We observe that for $\sigma = \emptyset$, the interpretations of none of the predicates of $\Sigma$ are required to be preserved. We can then easily see that the following lemma is true. Below cardinality of a $\Sigma-$structure refers to the cardinality of its universe.

**Lemma 2** *A sentence $\varphi \in \mathbf{EBS}_\Sigma(\emptyset)$ iff either (i) $\varphi$ is unsat or (ii) there exists a least finite cardinal $\mathcal{B}$ such that $\varphi$ has a model of (a) every cardinality $\geq \mathcal{B}$ if $\varphi$ has an infinite model and (b) every cardinality ranging from $\mathcal{B}$ to the cardinality of $\varphi$'s largest model if $\varphi$ has only finite models.*

*Proof*:
Suppose $\varphi \in \mathbf{EBS}_\Sigma(\emptyset)$. Let $\mathcal{B}$ be the bound for $\phi$. Suppose it has an infinite model. We show that there is a model of $\varphi$ of every cardinality $\geq \mathcal{B}$.
Take any finite cardinal $n \geq \mathcal{B}$. Since $\varphi$ has an infinite model, by Compactness Theorem, there cannot be any bound on the domain-size of the finite models of $\varphi$. Then consider a model $M$ of $\varphi$ of domain-size $\geq n$. Since $\varphi \notin \mathbf{EBS}_\Sigma(\emptyset)$, there exists $M_1 \subseteq M$ satisfying the $\mathbf{EBS}_\Sigma(\emptyset)$ conditions. Then take any $M_2$ such that $M_1 \subseteq M_2 \subseteq M$ and s.t. $|\mathcal{U}_{M_2}| = n$. Then there exists $M_2'$ such that $M_2'$ has the same domain as $M_2$ and $M_2' \models \varphi$. In other words, there is a model of $\varphi$ of size $n$.
Take any infinite cardinal. Since $\varphi$ has an infinite model, using the Upward and Downward Löwenheim-Skolem theorems, we conclude that there is a model of $\varphi$ of every infinite cardinality.
Summing up, $\varphi$ has models of every cardinality $\geq \mathcal{B}$.

Suppose $\varphi$ has only finite models. Clearly then it cannot have models of unbounded size because then by Compactness Theorem, it would have an infinite model. Then there exists a finite cardinal $N$ such that the domain-size of $\varphi'$s largest model is $N$. Then by a similar argument as above, we can show that $\varphi$ has models of all domain-sizes ranging from $\mathcal{B}$ to $N$.
Thus the 'only-if' part of the Lemma holds.

We now prove the 'if' part. If $\varphi$ is unsatisfiable, $\varphi \in \mathbf{EBS}_\Sigma(\emptyset)$ trivially. Suppose $\varphi$ is satisfiable and has an infinite model and there exists a smallest finite cardinal $\mathcal{B}$ such that $\varphi$ has a model of every cardinality $\geq \mathcal{B}$. Take any model $M$ of $\varphi$ of domain-size $\geq \mathcal{B}$. Consider any $M_1 \subseteq M$ such that $|\mathcal{U}_{M_1}| = \mathcal{B}$. Consider any $M_2$ such that $M_1 \subseteq M_2 \subseteq M$. Now from the premise, there exists a model $M_3$ of $\varphi$ of cardinality $|\mathcal{U}_{M_2}|$. Then consider the $\Sigma-$structure $M_2'$ which is isomorphic to $M_3$ and which has universe $\mathcal{U}_{M_2}$. Since the set of models of $\varphi$ is closed under isomorphism, $M_2' \models \varphi$ and is the desired model. Thus $\varphi \in \mathbf{EBS}_\Sigma(\emptyset)$.
If $\varphi$ is satisfiable, has only finite models and has a model of every cardinality ranging from $\mathcal{B}$ to the cardinality of $\varphi$'s largest model, then by a similar argument as the one above, we can show that $\varphi \in \mathbf{EBS}_\Sigma(\emptyset)$.
Thus the 'if' part holds as well proving the lemma. ■

Using the above lemma, we can see that many well-known classes of $\mathbf{FO}_\Sigma$ sentences fall within $\mathbf{EBS}_\Sigma(\emptyset)$. It can be shown that the following classes without functions but with corresponding 'function' predicates (i.e. we disallow functions but allow in their place predicates which represent the functions by including function axioms to be satisfied by the predicates) are also inside $\mathbf{EBS}_\Sigma(\emptyset)$.

1. The Löb-Gurevich class - $\mathbf{FO}_\Sigma$ without equality containing only unary predicates and only unary functions.

2. The Gurevich class - the set of all $\exists^*$ $\mathbf{FO}_\Sigma$ sentences possibly containing functions too.

We prove the inclusion of the above classes in $\mathbf{EBS}_\Sigma(\emptyset)$ below.

**Lemma 3** *The Löb-Gurevich Class* $\subseteq \mathbf{EBS}_\Sigma(\emptyset)$

*Proof:*
Consider the Löb-Gurevich Class. This is the set of all $\mathbf{FO}_\Sigma$ sentences without equality over a vocabulary which contains only unary predicates and only unary function symbols.

As shown in [2](pp. 251), every sentence $\phi$ in the Löb-Gurevich class is satisfiable over the same domains as a Löwenheim sentence $\psi$. Then $\phi$ and $\psi$ have the same spectra. But we know that if signature of $\psi$ is $\Sigma$, then $\psi \in \mathbf{EBS}_\Sigma(\Sigma) \subseteq \mathbf{EBS}_\Sigma(\emptyset)$. Then using Lemma 2 and the fact above that the spectrum of $\phi$ is the same as that of $\psi$, we have that $\phi \in \mathbf{EBS}_\Sigma(\emptyset)$.

It is easy to see that a sentence in the Löb-Gurevich class, with functions replaced by the corresponding function predicates satisfying the function axioms will also be in $\phi \in \mathbf{EBS}_\Sigma(\emptyset)$. ∎

**Lemma 4** *The Gurevich Class* $\subseteq \mathbf{EBS}_\Sigma(\emptyset)$

*Proof:*
Consider an $\mathbf{FO}$ sentence $\phi$ in the Gurevich class. Let $\phi = \exists^* \mathbf{z} \varphi(\mathbf{z})$ where $\varphi$ could contain function symbols and equality too in addition to predicates. Suppose $M \models \phi$ such that $|\mathcal{U}_M| \geq |\phi|$ where $|\phi|$ denotes the length of $\phi$. Then there exists elements $\mathbf{a}$ from $\mathcal{U}_M$ such that $M \models \varphi(\mathbf{a})$. Consider all the terms occuring in $\varphi(\mathbf{a})$. Let $\mathcal{T}$ be the set of all the values that these terms evaluate to in $M$. Note that $|\mathcal{T}| < |\phi|$. Choose an element $a^* \in \mathcal{U}_M \setminus \mathcal{T}$. Consider the model $M'$ such that $\mathcal{U}_{M'} = \mathcal{U}_M$ and $M'|_\mathcal{P} = M|_\mathcal{P}$ where $\mathcal{P}$ is the set of all predicate symbols of $\Sigma$. Now for a $k$−ary function symbol $f \in \Sigma$, denote the interpretations of $f$ in $M$ and $M'$ as $f_M$ and $f_{M'}$. Then $f_{M'}$ is defined as follows: Let $f(t^1, \ldots, t^k)$ be a term in $\varphi(\mathbf{a})$. Let $t^1_M, \ldots t^k_M$ be the evaluations of $t^1, \ldots, t^k$ in $M$ (then the former are values in $\mathcal{T}$). Then $f_{M'}(t^1_M, \ldots, t^k_M) = f_M(t^1_M, \ldots, t^k_M)$ (observe that this value must also be in $\mathcal{T}$). For all other values $a_1, \ldots, a_k$ for which there do not exist terms $t^1, \ldots, t^k$ such that (i) $t^i_M = a_i$ for $1 \leq i \leq k$ and (ii) $f(t^1, \ldots, t^k)$ is a term in $\varphi(\mathbf{a})$, assign $f(a_1, \ldots, a_k) = a^*$.

Then consider the substructure $M_1$ of $M'$ generated by $\mathcal{T} \cup \{a^*\}$. One can see that by our construction above $\mathcal{U}_{M_1} = \mathcal{T} \cup \{a^*\}$. Consider the substructure $M_2$ generated by any $\mathcal{U}$ s.t. $\mathcal{U}_{M_1} \subseteq \mathcal{U} \subseteq \mathcal{U}_M$. Again by our construction $\mathcal{U}_{M_2} = \mathcal{U}$. Then for all terms $t$ appearing in $\varphi(\mathbf{a})$, $t_{M_2} = t_{M'}$ (since $M_2 \subseteq M'$) and by construction $t_{M'} = t_M$. Then in $M_2$, all the terms evaluate to exactly the same values as they evaluate to in $M$. Now since $M_2|_\mathcal{P} = M'|_\mathcal{P} = M|_\mathcal{P}$ (the first equality is because $M_2 \subseteq M'$ and the second is by construction), $M_2 \models \varphi(\mathbf{a})$ i.e. $M_2 \models \phi$.

This thus shows that given a model $M$ of $\phi$ of size $\geq \mathcal{B} = |\phi|$, it is possible to construct a model $M'$ of $\phi$ by preserving the interpretations of all the predicates of $\Sigma$ and changing the interpretations of only the functions such that $M'$ has the extensible bounded submodel property.

It is easy to see that $\Phi$ which is obtained by replacing the functions in $\phi$ by their corresponding function predicates and adding the function axioms will also satisfy the above property.

As a corollary of this we get that if $\Phi$ has a model with domain-size $n \geq \mathcal{B}$, it has models of all domain-sizes ranging from $\mathcal{B}$ to $n$. Then if $\Phi$ has models of unbounded domain-size - which is iff it has an infinite model - then by the previous statement and the Upward and Downward Löwenheim-Skolem theorems, it has models of all domain-sizes $\geq \mathcal{B}$. Thus $\Phi \in \mathbf{EBS}_\Sigma(\emptyset)$ where $\Sigma$ is the vocabulary of $\Phi$. ∎

We have already seen earlier that $\mathbf{FO}_\Sigma[1]$ and $\mathbf{FO}_\Sigma[1]$ with counting are inside $\mathbf{EBS}_\Sigma(\Sigma)$ and hence inside $\mathbf{EBS}_\Sigma(\emptyset)$. We now show below that $\mathbf{FO}_\Sigma[2]$ ($\mathbf{FO}_\Sigma$ with only two variables) is contained in $\mathbf{EBS}_\Sigma(\emptyset)$.

**Lemma 5** $\mathbf{FO}_\Sigma[2] \subseteq \mathbf{EBS}_\Sigma(\emptyset)$

*Proof:*
We refer to the proof of the finite model property of $\mathbf{FO}_\Sigma[2]$ as shown in [5]. Firstly it is sufficent to consider only arity $\leq 2$ predicates when dealing with $\mathbf{FO}_\Sigma[2]$. As shown in [5], an $\mathbf{FO}_\Sigma[2]$ sentence $\phi$ over $\Sigma$ is equisatisfiable over the same domains with an $\mathbf{FO}_\Sigma[2]$ sentence $\psi$ over a vocabulary $\Sigma_1$ which contains no arity $> 2$ predicates. The paper further constructs a sentence $\theta$ over the same vocabulary $\Sigma_1$, in Scott Normal Form. i.e. $\theta$ is of the form
$\theta = \forall x \forall y \alpha(x, y) \wedge \bigwedge_{i=1}^{i=m} \forall x \exists y (\beta_i(x, y) \wedge x \neq y)$
where $\alpha$ and the $\beta_i$'s are quantifier-free. $\theta$ is such that $\psi$ has a model whose domain-size is $n \geq 2$ iff $\theta$ has a model with domain-size $n$. Then it is shown that if $\theta$ is satisfiable, it has a model with domain-size atmost $3m.2^n$ where $n = |\Sigma_1|$.

We now show that if $\theta$ is satisfiable, either it has only finitely many models or it has models with all domain-sizes $\geq 3m.2^n$. Then $\psi$ and hence $\phi$ would have either finitely many models or models with all domain-sizes $\geq 3m.2^n$. The argument is a simple extension of the procedure (given in [5]) used in constructing the small model for $\theta$.

Firstly if all models of $\theta$ have domain-size $\leq 2^n$ then all models of $\psi$ and hence $\phi$ would also have domain-size $\leq 2^n$. If not then consider a model of $\theta$ of domain size $> 2^n$. Then there exists an element in the model which is not a 'king' (See [5] for the definition of 'king'). Now from this model, construct the small model $M$ for $\theta$ as per the procedure. Then in $M$ consider an element $d_{ij}$ (using the same notation as in the paper). We introduce a new element $d'$ in $M$ which will 'mimic' $d_{ij}$. In particular, $d'$ has the same 1-type as $d_{ij}$. Next, for every element $e$ in $M$ different from $d_{ij}$ and $d'$, we make the 2-type of $(d', e)$ the same as the 2-type of $(d_{ij}, e)$. Note that this 2-type assignment cannot cause a conflict for the 1-type of $e$ or for the 1-type of $d'$. Further this assignment of the 2-type between $d'$ and $e$ cannot conflict with the 2-types of the already existing (i.e. prior to the introduction of $d'$) pairs of elements since this 2-type is being assigned for a completely new pair of elements. Finally assign the 2-type of $(d', d_{ij})$ to be the 2-type of any two elements of $M$ whose 1-types are the same as that of $d'$ and $d_{ij}$, for example say, $e_{ij}$ and $d_{ij}$. Thus there would be no conflicts at all in this new structure, call it $M^+$.

Now we show that $M^+ \models \theta$. While the existing elements already have 'skolem witnesses', it remains to choose skolem witnesses for $d'$. To provide skolem witnesses for $d'$, we will again just mimic what was done for $d_{ij}$ i.e. for each $k \in \{1, \ldots, m\}$, we will choose the same skolem witness for $d'$ for $k$ as the the skolem witness $g_k(d_{ij})$ for $d_{ij}$ for $k$. Now by construction, the 2-types of $(d_{ij}, g_k(d_{ij}))$ and $(d', g_k(d_{ij}))$ are the same. Since the 2-type of $(d_{ij}, g_k(d_{ij}))$ ensures that $\beta_k(d_{ij}, g_k(d_{ij}))$ was True, $\beta_k(d', g_k(d_{ij}))$ will also be True. Then the second conjunct above of $\theta$ would be true in $M^+$. Since the 2-type between $d'$ and any element $e$ of $M$ is one that was already existing in $M$ and since the first conjunct of $\theta$ was true in $M$, the first conjunct would be true in $M^+$ as well. Then $M^+ \models \theta$. Note that $|U_{M^+}| = |U_M| + 1$.

We can now use the same procedure on $M^+$ to get another model of $\theta$ which contains one more element than the number of elements in $M^+$ and so on. Thus $\theta$ has models with all domain-sizes $\geq 3m.2^n$ since the initial model $M$ had domain-size $\leq 3m.2^n$. Then back-tracking, $\psi$ and hence $\phi$ has models with all domain-sizes $\geq 3m.2^n$.

Thus $\phi$ is either unsatisfiable, or has finitely many models (all having domain-size $\leq 2^n$) or has models with all domain-sizes $\geq 3m.2^n$. Then by Lemma 2, $\phi \in \mathbf{EBS}_\Sigma(\emptyset)$ with $\mathcal{B} \leq 3m.2^n$. ∎

It can also be shown easily that the class of all $\mathbf{FO}_\Sigma$ sentences without equality which have the finite model property is contained inside $\mathbf{EBS}_\Sigma(\emptyset)$.

**Lemma 6** $\mathbf{FO}_\Sigma$ *without equality and with finite model property* $\subseteq \mathbf{EBS}_\Sigma(\emptyset)$

*Proof:*
It is well known that if an $\mathbf{FO}_\Sigma$ without equality sentence $\phi$ has a model of size $n$, then it also has a model of size $n + 1$. Then if $\phi$ has finite model property, then either $\phi$ is unsatisfiable or there exists a least $\mathcal{B}$ s.t. $\phi$ has models of all cardinalities $\geq \mathcal{B}$. Then from Lemma 2, $\phi \in \mathbf{EBS}_\Sigma(\emptyset)$. ∎

Any $\mathbf{EBS}_\Sigma(\emptyset)$ sentence (and hence one without equality) has the finite model property. This shows that

$$\mathbf{EBS}_\Sigma(\emptyset) = \mathbf{FO}_\Sigma \text{ without equality and with finite model property}$$

An example of an $\mathbf{FO}_\Sigma$ without equality subclass with finite model property is the Gödel-Kalmár-Schütte class over $\Sigma$ which is the set of all $\exists^* \forall \forall \exists^*$ $\mathbf{FO}_\Sigma$ sentences without equality. This class has the small model property (see [2]). Then this class is contained inside $\mathbf{EBS}_\Sigma(\emptyset)$.

## 5.2 An interesting property of $\mathbf{EBS}_\Sigma(\emptyset)$ without equality

Consider a sentence $\phi \in \mathbf{EBS}_\Sigma(\emptyset)$ without equality. Given a model $M$ of $\phi$ of domain-size $\geq \mathcal{B}$ where $\mathcal{B}$ is the bound for $\phi$, it is not necessary that $M$ have a bounded core all of whose extensions within $M$ also satisfy $\phi$. But interestingly, we show in this section that one can always get another model $M'$ with the same universe as that of $M$ such that $M'$ has a bounded core (of size $\leq \mathcal{B}$) all of whose extensions within $M'$ are also models of $\phi$.

Consider a satisfiable sentence $\phi \in \mathbf{EBS}_\Sigma(\emptyset)$ without equality. Then there exists a $\mathcal{B}$ for $\phi$ such that for all domain sizes $n \geq \mathcal{B}$ we can construct a model $M$ of domain-size $n$. Then there exists a model $\hat{M}$ of $\phi$ of domain-size $\mathcal{B}$. Then it is possible to construct an isomorphic model $M^*$ whose universe is $\mathcal{U}_{M^*} = \{a_1, \ldots, a_\mathcal{B}\} \subseteq \mathcal{U}_M$. Then $M^* \models \phi$. We will use this model to construct the desired $M'$. Let $f : \mathcal{U}_M \to \mathcal{U}_{M^*}$ be any function from $\mathcal{U}_M$ to $\mathcal{U}_{M^*}$ such that $f(a_i) = a_i$ for all $i \in \{1, \ldots, \mathcal{B}\}$. For each $k-$ary predicate $P \in \Sigma$, denote the interpretation of $P$ in $M'$ and $M^*$ as $P^{M'}$ and $P^{M^*}$ respectively. Then define $P$ in $M'$ such that $M' \models P(b_1, \ldots, b_k)$ iff $M^* \models P(f(b_1), \ldots, f(b_k))$ for $b_1, \ldots, b_k \in \mathcal{U}_M$. Let w.l.o.g., $\phi$ be given by

$$\phi = \forall \mathbf{z}_0 \exists v_1 \forall \mathbf{z}_1 \exists v_2 \forall \mathbf{z}_2 \ldots \exists v_r \forall \mathbf{z}_r \xi(\mathbf{z}, v_1 \ldots, v_r)$$

where $\mathbf{z} = (\mathbf{z}_0, \ldots, \mathbf{z}_r)$ and $\xi$ is quantifier-free and in CNF. We will show that $M' \models \phi$.

Consider an instantiation $\mathcal{Z}$ of $\mathbf{z}$ from $\mathcal{U}_{M'} = \mathcal{U}_M$. We need to choose values $d_1, \ldots, d_r$ for the variables $v_1, \ldots v_r$ such that $M' \models \xi(\mathcal{Z}, d_1, \ldots, d_r)$. Let $\mathcal{Z}^*$ be the vector of values obtained by subjecting $\mathcal{Z}$ to $f$ component-wise. Then $\mathcal{Z}^*$ is a vector of values from $\mathcal{U}_{M^*}$. Now since $M^* \models \phi$, there exist values $d_1^*, \ldots, d_r^*$ from $\mathcal{U}_{M^*}$ such that $M^* \models \xi(\mathcal{Z}^*, d_1^*, \ldots, d_r^*)$. Then for each $i \in \{1, \ldots, r\}$ choose $d_i = d_i^*$. Now consider an instance of $P$ in $\xi(\mathcal{Z}, d_1, \ldots, d_r)$. Say it is $P(e_1, \ldots, e_k)$. Then by construction above $M' \models P(e_1, \ldots, e_k)$ iff $M^* \models P(f(e_1), \ldots, f(e_k))$. But since $\mathcal{Z}^*$ is the image of $\mathcal{Z}$ under $f$ and $f(d_i) = f(d_i^*) = d_i^*$, we find that $P(f(e_1), \ldots, f(e_k))$ is the corresponding instance of $P(e_1, \ldots, e_k)$ in $\xi(\mathcal{Z}^*, d_1^*, \ldots, d_r^*)$. Then since the latter is True in $M^*$, $\xi(\mathcal{Z}, d_1, \ldots, d_r)$ is True in $M'$. Since $\mathcal{Z}$ was arbitrary, $M' \models \phi$.

Now let $M_1$ be the substructure of $M'$ generated by $\{a_1, \ldots, a_\mathcal{B}\}$. Let $M_1 \subseteq M_2 \subseteq M'$. Then since $M_2 \subseteq M'$, for each $k-$ary $P \in \Sigma$, $M_2 \models P(b_1, \ldots, b_k) \leftrightarrow M' \models P(b_1, \ldots, b_k) \leftrightarrow M^* \models P(f(b_1), \ldots, f(b_k))$. Then by a similar argument as the one shown above, we can infer that $M_2 \models \phi$. Thus $M'$ has the extensible bounded core property where the core has domain-size $\leq \mathcal{B}$.

Interestingly, in the above construction, we note that in making $\phi$ True in $M_2$, the 'inner' existential variables always take values from a fixed set of size $\mathcal{B}$, namely $\{a_1, \ldots, a_\mathcal{B}\}$. Then we construct the formula $\psi$ as given in the proof of Theorem 1 (in section 3) i.e.
$\psi = \exists x_1 \ldots \exists x_\mathcal{B} \forall \mathbf{z}_0 \forall \mathbf{z}_1 \ldots \forall \mathbf{z}_r\ (\chi)$ where
$\chi = (\bigvee_{u_1 \in \mathcal{V}} \ldots \bigvee_{u_r \in \mathcal{V}} \xi(\mathbf{z}, v_1 \mapsto u_1, \ldots, v_r \mapsto u_r))$
Then we can see that on account of the extensible bounded core property of $M'$, $M' \models \psi$. As shown above, for *any* model $M$ of domain-size $\geq \mathcal{B}$, we can construct a model $M'$ having the same domain as $M$ and which has the extensible bounded core property. Then the spectrum of $\phi$ is a subset of the spectrum of $\psi$. Clearly it is the case that if $M \models \psi$, then $M \models \phi$. Then the spectrum of $\psi$ is a subset of the spectrum of $\phi$. Then we conclude that $\phi$ and $\psi$ have the same spectrum. Thus we can answer any questions regarding the spectrum of $\phi$ by just asking the same questions of the spectrum of $\psi$ which is a $\mathbf{BSR}_\Sigma$ sentence.

While we do not know whether this property extends to the whole of $\mathbf{EBS}_\Sigma(\emptyset)$ with equality, in some cases it does. Consider a sentence $\phi$ in the Gurevich class seen above. As shown in the proof of Lemma 4, for every model $M$ of $\phi$, as above, one can construct $M'$ by changing only the interpretations of the 'function' predicates (and by keeping the interpretations of the other predicates as in $M$) such that $M'$ has the above mentioned extensible bounded core property.

## 5.3 The SAT problem for the $\mathbf{EBS}_\Sigma(\emptyset)$ class

**Lemma 7** *If $\varphi \in \mathbf{EBS}_\Sigma(\emptyset)$, the following problems are decidable:*
*(i) Is $\varphi$ satisfiable?*
*(ii) Does $\varphi$ have a finite model?*

*Proof:* By Lemma 2, if $\varphi \in \mathcal{EBS}$, it is either unsatisfiable or has a finite model of size atmost $\mathcal{B}$, where $\mathcal{B}$ is the finite cardinal in Lemma 2. By Herbrand's theorem, if $\varphi$ is unsatisfiable, there must exist a finite set of ground clauses of $\varphi$ that is propositionally unsatisfiable. We can therefore construct a decision procedure by interleaving steps of the following:

- A procedure that recursively enumerates the ground clauses in the Herbrand Universe of $\varphi$ and checks the propositional unsatisfiability of finite subsets of ground clauses, and

- A procedure that recursively enumerates finite models of $\varphi$ and checks whether $\varphi$ is true in the model.

Since $\varphi$ is either unsatisfiable or has a finite model of size at most $\mathcal{B}$, one of these is guaranteed to terminate. If we terminate by detecting unsatisifability of $\varphi$, then $\varphi$ has no model. Otherwise, we terminate with a finite model of $\varphi$. This gives us a decision procedure that serves to check satisfiability of $\varphi$ and also detects if $\varphi$ has a finite model. ∎

Now $\mathbf{BSR}_\Sigma \subseteq \mathbf{EBS}_\Sigma(\Sigma) \subseteq \mathbf{EBS}_\Sigma(\sigma)$ for $\sigma \subseteq \Sigma$. Since the SAT problem for $\mathbf{BSR}_\Sigma$ is NEXPTIME-complete, the SAT problem for $\mathbf{EBS}_\Sigma(\sigma)$ for any $\sigma \subseteq \Sigma$ (and in particular $\sigma = \emptyset$), is atleast NEXPTIME-hard.
For some special subclasses of $\mathbf{EBS}_\Sigma(\emptyset)$, we identify below the complexity of the SAT problem for these subclasses.

**Lemma 8** *Assume a sub-class $\mathcal{S}$ of $\mathbf{EBS}_\Sigma(\emptyset)$ for which $\mathcal{B}$ for every sentence $\phi$ of the sub-class is atmost $f(\phi)$ where $f : \mathcal{S} \to \mathbb{N}$ is an efficiently (polynomial time) computable function. Then SAT-$\mathcal{S} \in \text{NTIME}(f(\phi)^{O(|\phi|)})$ where SAT-$\mathcal{S}$ denotes the SAT problem for $\mathcal{S}$ and $|\phi|$ denotes the length of $\phi$.*

*Proof*:
Given a formula $\phi \in \mathcal{S} \subseteq \mathbf{EBS}_\Sigma(\emptyset)$, we construct a Turing machine which does the following:

1. It first brings $\phi$ to its PCNF $\varphi$. The time taken to do so is $O(|\phi|)$. The length of $\varphi$, i.e. $|\varphi| = O(|\phi|)$.

2. It computes $\mathcal{B} = f(\phi)$. The time taken for this is $O(q(|\phi|))$ where $q$ is some polynomial.

3. It then guesses a number $n$ between 0 and $\mathcal{B}$. It then constructs the set $\{1, \ldots, n\}$ which would serve as the universe $\mathcal{U}_M$ of a $\Sigma-$structure $M$. The time required to construct the universe is $O(n) = O(\mathcal{B})$.

4. It then guesses the $\Sigma-$structure $M$. Let $r$ be the number of predicates in $\phi$ ($r = O(|\phi|)$). Let $s$ be the maximum arity of these predicates ($s = O(|\phi|)$). To interpret any predicate $P$ requires specifying the truth value of $P$ for each $s-$tuple of arguments from $\mathcal{U}_M$. Then the time required to guess an interpretation of $P$ is $O(n^s)$. Doing so for all predicates requires time $O(r \cdot n^s) = O(|\phi| \cdot n^{|\phi|})$. Thus the time to guess the $\Sigma-$structure $M$ is $O(|\phi| \cdot n^{|\phi|})$. This also is the size of $M$ which we denote by $\| M \|$ i.e. $\| M \| = O(|\phi| \cdot n^{|\phi|})$.

5. It then checks whether $M \models \varphi$. It does so by enlisting out all the matrices obtained by instantiating the variables in the matrix of $\varphi$ with values from $\mathcal{U}_M$ and then checking whether these instantiated matrices evaluate to True in $M$. Given an instantiated matrix, the time required to check whether the matrix is True in $M$ is $O(\|M\| \cdot |\varphi|) = O(\|M\| \cdot |\phi|)$. If the length of the quantifier prefix of $\varphi$ is $p(= O(|\varphi|) = O(|\phi|))$, then the number of instantiated matrices is $O(n^p)$. Then the total time to check whether $M \models \varphi$ is $O(n^p) \cdot O(\|M\| \cdot |\phi|) = O(n^{|\phi|} \cdot |\phi| \cdot n^{|\phi|} \cdot |\phi|) = O(n^{2 \cdot |\phi|} \cdot |\phi|^2)$.

If $M \models \varphi$, then the TM returns Yes (i.e. $\varphi$ and hence $\phi$ is satisfiable and also would have constructed the model $M$ satisfying it). If all computations of $M$ fail, then it indeed must be that $\varphi$ and hence $\phi$ is unsatisfiable due to the bounded model property of $\phi$. Thus $M$ decides the satisfiability of $\phi$.

The time taken for any computation =
Time to compute $\varphi$ (A) + Time to compute $f$ (B) + Time taken to guess $n$ (C) + Time taken to guess $M$ with universe of size $n$ (D) + Time taken to check if $M \models \varphi$ (E)

Now (C) + (D) + (E)
$= O(n) + O(|\phi| \cdot n^{|\phi|}) + O(n^{2 \cdot |\phi|} \cdot |\phi|^2)$
$= O(n^{2 \cdot |\phi|} \cdot |\phi|^2)$
$= O(n^{O(|\phi|)})$
Then (A) + (B) + (C) + (D) + (E)
$= O(|\phi|) + O(q(|\phi|)) + O(n^{O(|\phi|)})$
$= O(n^{O(|\phi|)})$
$= O(f(\phi)^{O(|\phi|)})$
Thus SAT-$\mathcal{S} \in \text{NTIME}(f(\phi)^{O(|\phi|)})$. ∎

We get the following immediately as a consequence of the above Lemma.

**Corollary 2** *If for the sub-class $\mathcal{S}$ satisfying the conditions of Lemma 8, $f(\phi) \leq 2^{p(|\phi|)}$ for some polynomial $p$, then SAT-$\mathcal{S} \in \text{NEXPTIME}$. If $\mathcal{S}$ contains $\mathbf{BSR}_\Sigma$, then SAT-$\mathcal{S}$ is NEXPTIME-complete.*

We shall look at one such class as stated in the above lemma later in section 8).

# 6 Characterizing the spectra of the $\mathbf{EBS}_\Sigma(\sigma)$ classes

The *spectrum* of a sentence $\phi$ is defined as the set of cardinalities of all the finite models of $\phi$. The spectrum of $\phi$ is thus a subset of $\mathbb{N}$. It was proved by Ramsey [2] that the spectrum of any sentence in $\mathbf{BSR}_\Sigma$ is either finite or co-finite. By Theorem 1 (in section 3), since $\mathbf{BSR}_\Sigma$ is the syntactic characterization of $\mathbf{EBS}_\Sigma(\Sigma)$, it follows that the spectrum of any $\mathbf{EBS}_\Sigma(\Sigma)$ sentence is also finite or co-finite.

Let $\mathcal{F}_\Sigma$ be the set of all $\mathbf{FO}_\Sigma$ sentences which have finite or co-finite spectra. Then from above, $\mathbf{EBS}_\Sigma(\Sigma) \subseteq \mathcal{F}_\Sigma$. We now ask: how does $\mathbf{EBS}_\Sigma(\sigma)$ compare with $\mathcal{F}_\Sigma$?

Lemma 2 (in section 5.1) immediately shows that $\mathbf{EBS}_\Sigma(\emptyset) = \mathcal{F}_\Sigma$.
Thus for any $\Sigma$, $\mathbf{EBS}_\Sigma(\emptyset)$ is exactly the set of all $\mathbf{FO}_\Sigma$ sentences which have finite or co-finite spectra.
From Theorem 2 (in section 4.1), this implies that unless $\Sigma$ contains only monadic predicates, for each non-empty $\sigma \subseteq \Sigma$,

$$\mathbf{EBS}_\Sigma(\sigma) \subsetneq \mathcal{F}_\Sigma$$

If $\Sigma$ contains only monadic predicates, we have

$$\mathbf{EBS}_\Sigma(\Sigma) = \mathcal{F}_\Sigma$$

This then answers the question posed above. $\mathcal{F}_\Sigma$ is exactly $\mathbf{EBS}_\Sigma(\sigma)$ if $\Sigma$ has only monadic predicates. If $\Sigma$ contains atleast one predicate of arity $\geq 2$, then $\mathbf{EBS}_\Sigma(\sigma)$ is a strict semantic subset of $\mathcal{F}_\Sigma$ for a non-empty $\sigma \subseteq \Sigma$.
Let for a set $A$ of sentences, $\mathcal{S}(A)$ denote the set of spectra of the sentences in $A$. We then ask the question: How do $\mathcal{S}(\mathbf{EBS}_\Sigma(\sigma))$ for different $\sigma$ compare?
We have the following lemma.

**Lemma 9** *Consider a set $\mathsf{S}$ which is finite or co-finite. Then there exists a $\mathbf{BSR}_\emptyset$ formula whose spectrum is $\mathsf{S}$.*

*Proof*:
Suppose $\mathsf{S}$ is finite. Let $\mathsf{S} = \{k_1, \ldots, k_n\}$. For each $k_i$, construct $\phi_i$ as follows:

$$\phi_i = \exists x_1 \ldots \exists x_{k_i} \forall y ((\bigwedge_{j=1}^{j=k_i} \bigwedge_{l=j+1}^{l=k_i} (x_j \neq x_l)) \wedge \bigvee_{j=1}^{j=k_i} (y = x_j))$$

$\phi_i$ asserts that there are exactly $k_i$ elements in any model of $\phi_i$. Then the formula

$$\phi_\mathsf{S} = \bigvee_{i=1}^{i=n} \phi_i$$

has a spectrum which is exactly $\mathsf{S}$. Note that since $\phi_\mathsf{S}$ is a disjunction of $\mathbf{BSR}_\emptyset$ sentences, $\phi_\mathsf{S} \in \mathbf{BSR}_\emptyset$.
Suppose $\mathsf{S}$ is co-finite. Then let
$\mathsf{S} = \{k_1, k_2, \ldots, k_n\} \cup \{\mathcal{B} + i \mid i \geq 0\}$ for some $n > 0$ where $\mathcal{B}$ is larger than all the $k_i$s.
For each $k_i$, we construct $\phi_i$ as shown above. We also construct the following sentence $\phi_\mathcal{B}$

$$\phi_\mathcal{B} = \exists x_1 \ldots \exists x_\mathcal{B} (\bigwedge_{j=1}^{j=\mathcal{B}} \bigwedge_{l=j+1}^{l=\mathcal{B}} (x_j \neq x_l))$$

Then the spectrum of $\phi_\mathcal{B}$ is exactly $\{\mathcal{B} + i \mid i \geq 0\}$. Then the formula

$$\phi_\mathsf{S} = (\bigvee_{i=1}^{i=n} \phi_i) \vee \phi_\mathcal{B}$$

has spectrum which is exactly $\mathsf{S}$. Again since $\phi_\mathsf{S}$ is a disjunction of $\mathbf{BSR}_\emptyset$ sentences, $\phi_\mathsf{S} \in \mathbf{BSR}_\Sigma$. ∎

**Corollary 3** *Consider a set S which is finite or co-finite. Then for every $\Sigma$, there exists a $\mathbf{BSR}_\Sigma$ formula whose spectrum is S.*

*Proof*:
By Lemma 9, there exists a $\mathbf{BSR}_\emptyset$ formula $\phi_S$ whose spectrum is S. Then consider the formula $\psi_S$ given by

$$\psi_S = \phi_S \wedge \bigwedge_{Q \in \Sigma} \forall \mathbf{z}_Q Q(\mathbf{z}_Q)$$

where $\mathbf{z}_Q$ is a vector of variables of length equal to the arity of $Q$. Any finite model of $\psi_S$ is also a model of $\phi_S$ and hence its cardinality must be in S. Conversely, for any $k \in$ S, there is a model of $\phi_S$ whose cardinality is $k$. Since $\phi_S$ is over the empty vocabulary, this model is just a universe of size $k$. Then this model can be expanded to a model of $\psi_S$ by interpreting all the predicates of $\Sigma$ as 'fully' True. Thus the spectrum of $\psi_S$ is exactly S. ∎

From the above, we see that for any $\Sigma$, $\mathcal{S}(\mathbf{BSR}_\Sigma)$ is exactly the set of all finite and co-finite sets. Then since from Theorems 1 (section 3) and 2(1) (section 4.1), $\mathbf{BSR}_\Sigma \subseteq \mathbf{EBS}_\Sigma(\Sigma) \subseteq \mathbf{EBS}_\Sigma(\sigma) \subseteq \mathcal{F}_\Sigma$, we have that $\mathcal{S}(\mathbf{EBS}_\Sigma(\sigma))$ is exactly the set of all finite and co-finite sets. Thus, for all $\Sigma$ and all $\sigma \subseteq \Sigma$, we have a complete characterization of the spectra of $\mathbf{EBS}_\Sigma(\sigma)$ as the set of all finite and co-finite sets.
Further since any $\mathbf{EBS}_\emptyset(\emptyset)$ sentence $\phi$ can be easily extended to a sentence of $\mathbf{EBS}_\Sigma(\sigma)$ which has the same spectrum as $\phi$, the $\mathbf{EBS}_\emptyset(\emptyset)$ class can be seen as the 'minimum' class in the $\mathbf{EBS}$ heirarchy whose set of all spectra is exactly the set of all finite and co-finite sets.

## 7 Undecidability of the $\mathbf{EBS}_\Sigma(\sigma)$ classes

While the satisfiability problem for $\mathbf{EBS}_\Sigma(\sigma)$ is decidable, the membership problem for $\mathbf{EBS}_\Sigma(\sigma)$ is not always decidable. Below we look at the cases where $\mathbf{EBS}_\Sigma(\sigma)$ is decidable and where it is undecidable. In the latter case, we show a reduction from the halting problem.
As shown in [7], while the set of all those TMs which halt on the empty tape is undecidable, researchers set out to find out subsets of this set, which are undecidable. For any pair $(s, l)$ of natural numbers, define the set $\mathcal{A}_{(s,l)}$ to be the set of all the TMs having $s + 1$ states (including the halting state) and $l$ symbols (including the blank symbol). Then it can be shown ([7]) that if $\mathcal{A}_{(s,l)}$ is undecidable, then $\mathcal{A}_{(x,y)}$ is undecidable for $s \leq x$ and $l \leq y$. A set $R$ of seven such pairs $(s, l)$ have been identified ([7]) such that $\mathcal{A}_{(s,l)}$ is undecidable and further these are such that there remain only a fixed finite number of pairs $(x, y)$ for which the decidability status of $\mathcal{A}_{(x,y)}$ is unknown. We use this set $R$ in the theorem below.

**Theorem 3** *For any $\sigma \subseteq \Sigma$,*

1. *If $\Sigma$ contains only unary predicates, $\mathbf{EBS}_\Sigma(\sigma)$ is decidable.*

2. *If $\Sigma$ contains at least $s+1$ unary predicates and atleast $l + 1$ predicates of arity $\geq 2$, where $(s, l) \in R$, then $\mathbf{EBS}_\Sigma(\sigma)$ is undecidable.*

*Proof:*
Statement (1) above is easy to see. If $\Sigma$ contains only unary predicates, then as seen earlier, $\mathbf{EBS}_\Sigma(\sigma) = \mathbf{EBS}_\Sigma(\Sigma) = \mathbf{FO}_\Sigma$. Then checking membership is trivial.
We now consider statement (2) above. Consider the pair $(s, l)$. Then as mentioned above, $\mathcal{A}_{(s,l)}$ is undecidable. We show a reduction from $\mathcal{A}_{(s,l)}$ to the set $\overline{\mathbf{EBS}_\Sigma(\sigma)}$ which is the complement of the set $\mathbf{EBS}_\Sigma(\sigma)$. Clearly $\overline{\mathbf{EBS}_\Sigma(\sigma)}$ is decidable iff $\mathbf{EBS}_\Sigma(\sigma)$ is decidable.
Consider $\mathcal{M}_1 \in \mathcal{A}_{(s,l)}$. Let $\mathcal{M}_1 = (Q, \Sigma, \Delta, \delta_1, q_0, \{h\})$ be a deterministic Turing Machine (DTM) with a two-way infinite tape, where $Q$ is the set of states such that $|Q| = s+1$, $\Sigma$ represents the input alphabet, $\Delta$ represents the tape alphabet such that $|\Delta| = l$, $\delta_1 : Q \times \Delta \to Q \times \Delta \times \{L, R\}$ represents the deterministic transition function, $q_0$ represents the starting state and $h$ represents the unique halting state.
We first modify $\mathcal{M}_1$ to a new DTM $\mathcal{M} = (Q \cup \{t\}, \Sigma, \Delta, \delta, q_0, \emptyset)$, where $t \notin Q$ and
$\delta(q, i) = \delta_1(q, i), \forall q \in Q \setminus \{h\}, \forall i \in \Delta;$
$\delta(h, i) = (t, i, R), \quad \delta(t, i) = (h, i, L) \quad \forall i \in \Delta$

Thus $\mathcal{M}$ mimics the behaviour of $\mathcal{M}_1$ exactly except when $\mathcal{M}_1$ reaches the halting state $h$. On reaching this state, $\mathcal{M}$ loops between states $h$ and $t$ forever. Note that $\mathcal{M}$ does not have a halting state.
Next, we construct a sentence $\Phi_\mathcal{M} \in \mathbf{FO}_\Sigma$ such that models of $\Phi_\mathcal{M}$ represent computations of $\mathcal{M}$ on the empty input tape that end in state $t$. For this, we proceed exactly as in the proof of Trakhtenbrot's theorem as presented in [6, pp 165-168] (details are therefore omitted here). We make the following slight changes though.

1. The proof in [6] uses only binary predicates whereas the arity $\geq 2$ predicates of $\Sigma$ need not necessarily be binary. But it is easy to see that binary predicates can always be simulated using higher arity predicates (as was shown earlier in section 4.1).

2. The proof in [6] introduces a binary predicate $H_q(p, n)$ for each state $q$ where $H_q(p, n)$ denotes that at time $n$, $\mathcal{M}$ is in state $q$ and its head is at position $p$ on the tape. We instead separate the state and the head-position information by using a unary predicate $U_q(n)$ for each state $q \in Q$ and a arity $\geq 2$ predicate $H$. $H(p, n, \ldots, n)$ is True iff the head is at position $p$ at time $n$. $U_q(n)$ denotes that $\mathcal{M}$ is in state $q$ at time $n$. If for a given $n$, all $U_q(n)$'s are False, it would be taken to mean that at time $n$, $\mathcal{M}$ is state $t$. The constraints can easily be recast in terms of these predicates. The total number of unary predicates needed above is $s + 1$ - which is possible to use from $\Sigma$ since $\Sigma$ contains atleast $s + 1$ unary predicates.

3. The proof in [6] uses the constant <u>min</u>. We get rid of this constant and replace all its occurences with a variable say $x$ which will be existentially quantified. Since $\mathcal{M}$ uses a two-way infinite tape, we remove the constraint that $x$ is the minimum element. $x$ would just be some element which would represent the start time.

4. The proof in [6] considers the tape alphabet to be $\{0.1\}$. But the same constraints as for this tape alphabet, can be written for any general tape alphabet, in particular $\Delta$. Then one binary predicate $P_a(p, n)$ would be used for each $a \in \Delta$ with the constraint that for any $(p, n)$ exactly one these predicates is True. That requires $l$ binary predicates for all symbols together. But we use only $l - 1$ arity $\geq 2$ predicates - one predicate $P_a$ for each symbol in $\Sigma$ and assert that *atmost* one such predicate is True for each $(p, n)$. The case when all the $P_a$'s are False will be taken to mean that the blank symbol is present at position $p$ at time $n$. The total number of arity $\geq 2$ predicates needed would be $l - 1$ (for the $P_a$ predicates) + 1 (for $H$) + 1 (for the linear order) = $l + 1$. These can be used from $\Sigma$ since $\Sigma$ contains atleast $l + 1$ arity $\geq 2$ predicates.

5. The proof in [6] assumes a complete DTM. Here we do not require it to be so. Hence we need to capture the condition that $\mathcal{M}$ might halt on a state $q$ - which is not the halting state - just because there is no transition out from the state on the symbol $a$ currently under the head (Clearly the input is not accepted by this halting). Then we add the following constraint $\phi$ in $\Phi_1$ for each such $q$ and $a$:

   $\text{last}(y) = \forall x((x \neq y) \rightarrow (x < y))$

   $\phi(y) = \forall n \forall p((U_q(n) \land H(p, n, \ldots, n) \land P_a(p, n, \ldots, n)) \rightarrow \bigvee_{q \in Q} U_q(\text{last}(y)))$

   where $P_a(p, n, \ldots, n)$ is True iff at time $n$, the symbol at position $p$ on the tape is $a$. $\phi$ then asserts that if $q$ and $a$ are as mentioned above, the last state of the computation is not $t$.

6. The last condition is changed to assert that the last state of the computation is $t$

   $\Phi_2 = \exists x(\text{last}(x) \land \bigwedge_{q \in Q} \neg U_q(x))$.

7. For all those $P \in \Sigma$ which were not used in the constraints above, we just add the constraint $\psi_P = \forall \mathbf{z}_P P(\mathbf{z}_P)$ where $\mathbf{z}_P$ is a vector of length equal to the arity of $P$. These additional constraints are just to make $\Phi_\mathcal{M}$ have signature equal to $\Sigma$.

The above constructed $\Phi_\mathcal{M}$ can be seen to be in $\mathbf{FO}_\Sigma$. We now observe that.

A) A $\Sigma$−structure $M$ with $|\mathcal{U}_M| = \alpha$ for some cardinal $\alpha$ is a model of $\Phi_\mathcal{M}$ iff $M$ represents a computation of length $\alpha$ of the DTM $\mathcal{M}$ (on the empty input tape) which ends in state $t$.

We now show that $\mathcal{M}_1$ halts on the empty input tape iff $\Phi_\mathcal{M} \in \overline{\mathbf{EBS}_\Sigma(\sigma)}$.
*If part:* Suppose $\mathcal{M}_1$ does not halt at the halting state $h$. Then $\mathcal{M}$ either halts at a non-halting state or it has an infinite computation that does not through $t$. In either case, no computation of it ends in $t$. Since by (A) above, there is a 1-1 correspondance between computations of $\mathcal{M}$ that end in state $t$ and the models of $\phi_\mathcal{M}$, we find that $\phi_\mathcal{M}$ is <u>unsatisfiable</u>. Then $\phi_\mathcal{M} \in \mathbf{EBS}_\Sigma(\sigma)$ trivially. So $\phi_\mathcal{M} \notin \overline{\mathbf{EBS}_\Sigma(\sigma)}$.
*Only if part:* Suppose that $\mathcal{M}_1$ halts at the halting state $h$. Let $\mathcal{N}$ be the length of the 'halting' computation of $\mathcal{M}_1$. Then $\mathcal{M}$ reaches $h$ for the first time after $\mathcal{N}$ steps. We have the following two observations:

1. Consider the computation of $\mathcal{M}$ of length $\mathcal{N}+2k$ ($k \geq 0$) (since $\mathcal{M}$ is a DTM, there is only one computation of length $\mathcal{N} + 2k$ for each $k$). By the construction of $\mathcal{M}$, every such computation ends in state $h$ (and hence not at $t$). Then from observation (A) above, $\Phi_\mathcal{M}$ has no model of size $\mathcal{N} + 2k$ for all $k \geq 0$.

2. Consider the computation of $\mathcal{M}$ of length $\mathcal{N} + 2k + 1$ ($k \geq 0$). By construction, the computation ends in state $t$. Then by (A) above, there exists a model of $\Phi_M$ of size $\mathcal{N} + 2k + 1$, for every $k \geq 0$.

Suppose that $\Phi_\mathcal{M} \in \mathbf{EBS}_\Sigma(\emptyset)$. From observation (2), we know that $\Phi_\mathcal{M}$ has infinitely many models. Then by Lemma 2 (in section 5.1), there exists a finite cardinal $\mathcal{B}$ such that $\Phi_\mathcal{M}$ has a model of every size larger than $\mathcal{B}$. However, from observation (1), we know that by taking any $k$ such that $\mathcal{N} + 2k > \mathcal{B}$, $\Phi_M$ has no model of size $\mathcal{N} + 2k$. Therefore, $\Phi_\mathcal{M} \notin \mathbf{EBS}_\Sigma(\emptyset)$. Then since $\mathbf{EBS}_\Sigma(\sigma) \subseteq \mathbf{EBS}_\Sigma(\emptyset)$ we conclude that $\Phi_\mathcal{M} \in \overline{\mathbf{EBS}_\Sigma(\sigma)}$.
Thus $\mathcal{M}_1$ halts iff $\Phi_M \in \overline{\mathbf{EBS}_\Sigma(\sigma)}$. Thus $\overline{\mathbf{EBS}_\Sigma(\sigma)}$ is undecidable. Hence $\mathbf{EBS}_\Sigma(\sigma)$ is undecidable. ∎

The above proof ofcourse leaves out the $\Sigma$s which contain $s + 1$ unary predicates and $l + 1$ predicates of arity $\geq 2$ where the decidability status of $\mathcal{A}_{(s,l)}$ is unknown. We do not have an answer yet for the decidability of $\mathbf{EBS}_\Sigma(\sigma)$ for these $\Sigma$s but we conjecture that these must be undecidable as well.

While for the Σs considered in Theorem 3(2), the latter says that $\mathbf{EBS}_\Sigma(\sigma)$ is undecidable, it does not tell us whether these are r.e. or co-r.e. Indeed we do not have an answer yet to this question as well for all $\sigma$. For the case of $\sigma = \Sigma$, the $\mathbf{EBS}_\Sigma(\Sigma)$ class has a syntactic characterization in the form of $\mathbf{BSR}_\Sigma$. Then as seen in section 3, $\mathbf{EBS}_\Sigma(\Sigma)$ is r.e. and not co-r.e. If likewise, for the other $\sigma$'s too, we can find a syntactic characterization for them or in general a characterization in terms of any recursive class of formulae, then we know that they also would be r.e. and not co-r.e.

## 8 A Syntactic sub-class of $\mathbf{EBS}_\Sigma(\sigma)$

$\mathbf{BSR}_\Sigma$ gave a good starting point as a syntactic characterization of $\mathbf{EBS}_\Sigma(\Sigma)$. In this section, we attempt to provide a syntactic characterization of a fragment of the $\mathbf{EBS}_\Sigma(\sigma)$ class. Without loss of generality, let $\varphi$ be an $\mathbf{FO}_\Sigma$ sentence in prenex conjunctive normal form (PCNF). As usual, we consider $\Sigma$ as a relational vocabulary. Let $V(\varphi)$, $EV(\varphi)$ and $AV(\varphi)$ denote respectively the set of the leftmost existential, non-leftmost (or 'inner') existential and universal variables of $\varphi$. We introduce the following classification of predicates and their arguments in $\varphi$.

1. The $i^{th}$ argument (call it $x$) of an instance of predicate $P$ in $\varphi$ is called (a) *free* if $x \in V(\varphi)$ (b) *universal* if $x \in AV(\varphi)$, and (c) *existential* if $x \in EV(\varphi)$.

2. The *free support set* of an instance $I$ of $P$ in $\varphi$ is the set of all variables in $V(\varphi)$ that appear as arguments of $I$. Likewise, the *existential and universal support sets* of an instance $I$ of $P$ in $\varphi$ are the sets of variables in $EV(\varphi)$ and $AV(\varphi)$ respectively that appear as arguments of $I$.

3. An instance $I$ of predicate $P$ in $\varphi$ is called (a) *free* if its existential and universal support sets are both empty, (b) *universal* if its existential support set is empty and universal support set is non-empty, and (c) *existential* otherwise, i.e., if its existential support set is non-empty. Note that an existential instance of $P$ may have a universal argument.

4. Predicate $P$ in $\varphi$ is called (a) *free* if every instance of $P$ in $\varphi$ is free, (b) *universal* if every instance of $P$ in $\varphi$ is either universal or free and atleast one instance is universal, (c) *existential* if atleast one instance of $P$ in $\varphi$ is existential. Note that an existential predicate may have a universal or free instance.

5. Two instances of predicate $P$ in $\varphi$ are said to be *existentially distinguishable* with respect to variable $v$ if there is an integer $i$ such that (i) the $i^{th}$ argument of each instance is either free or existential (i.e. not universal) and (ii) $v$ appears as the $i^{th}$ argument of one instance but not as the $i^{th}$ argument of the other instance (thus $v \in EV(\varphi) \cup V(\varphi)$).

6. An instance $I$ of a predicate $P$ is said to have $+ve$ polarity in $\varphi$ (or $I$ is $+ve$ in $\varphi$) if it appears un-negated in $\varphi$. It is said to have $-ve$ polarity in $\varphi$ (or $I$ is $-ve$ in $\varphi$) if it appears negated in $\varphi$.

As an example, consider the sentence
$\varphi = \exists y \exists u \forall v \exists w \quad (P(y,y) \vee \neg Q(u,y) \vee R(y,v)) \wedge$
$\qquad\qquad (Q(v,u) \vee P(y,u) \wedge \neg R(w,v)))$
Here $V(\varphi) = \{y, u\}$, $AV(\varphi) = \{v\}$ and $EV(\varphi) = \{w\}$. For $P$, the free, universal and existential support sets of the first instance are $\{y\}, \emptyset$ and $\emptyset$ respectively, while those of the second instance are $\{y, u\}, \emptyset$ and $\emptyset$ respectively. Both instances are free and hence predicate $P$ is free. The first instance of $Q$ is free while the second instance is universal as its universal support set is $\{v\} \neq \emptyset$ and its existential support set is $\emptyset$. Hence $Q$ is universal. The first instance of $R$ is universal while its second instance is existential since its existential support set is $\{w\} \neq \emptyset$. Hence $R$ is existential. Also the two instances of $R$ are existentially distinguishable w.r.t. $w$ since (i) the first argument of each instance is not universal and (ii) $w$ appears as the first argument of the second instance but not as the first argument of the first instance. Finally, the first instance of $R$ is $+ve$ i.e. has $+ve$ polarity while the second instance of $R$ is $-ve$ i.e. has $-ve$ polarity in $\varphi$.

Let $\mathsf{U}$ denote the set of all unary predicates of $\Sigma$. Let $\mathcal{F}, \mathcal{A}$ denote respectively the set of all free and universal predicates of $\Sigma$. Let $E_\mathsf{U}$ denote the set of all variables in $EV(\varphi)$ which appear in $\varphi$ as the argument of some unary predicate. Let $\overline{E_\mathsf{U}} = EV(\varphi) \setminus E_\mathsf{U}$.

**Definition 2 (The $\mathbf{EDP}_\Sigma(\sigma)$ class)** *Let $\varphi$ be a $\mathbf{FO}_\Sigma$ sentence in PCNF. Then $\varphi$ is said to have **Existentially Distinguishable Predicates** preserving $\sigma \subseteq \Sigma$ if the following hold:*

1. *$\sigma \subseteq \mathsf{U} \cup \mathcal{F} \cup \mathcal{A}$*

2. *Equality (if present) $\in \mathcal{F} \cup \mathcal{A}$.*

3. *For any existential predicate $P$ of arity $\geq 2$, every pair of instances of $P$ appearing in $\varphi$ in different clauses and with different polarities, must be existentially distinguishable with respect to atleast one existential variable outside $E_\mathsf{U}$.*

*The set of all $\mathbf{FO}_\Sigma$ sentences having existentially distinguishable predicates preserving $\sigma$ is called the $\mathbf{EDP}_\Sigma(\sigma)$ class.*

Note that a unary predicate or a predicate with arity $\geq 2$ that is free or universal in $\varphi$ can be either in $\sigma$

or in $\Sigma \setminus \sigma$. As an example, consider the sentence
$\varphi = \exists x \exists y \forall z \exists v ((P(x,z) \lor R(y,z)) \land$
$(\neg P(v,y) \lor P(z,y)) \land (\neg R(x,z) \lor (z=x)))$
Here, $\Sigma = \{P, R\}$, $\mathsf{U} = \mathcal{F} = \emptyset$, $\mathcal{A} = \{R, =\}$ and $E_\mathsf{U} = \emptyset$. Consider $\sigma \subseteq \Sigma$ such that $P \notin \sigma$. Then we can see that conditions 1 and 2 in the **EDP** definition are satisfied. Predicate $P$ is existential as its second instance $P(v,y)$ is existential. For the first and second instances of $P$ from the left, the second instance is existential and lies in a different clause and with different polarity w.r.t. the first instance. Indeed then $v \notin E_\mathsf{U}$ existentially distinguishes the two instances. The second and third instances of $P$ have different polarities but lie in the same clause. The first and third instances of $P$ lie in different clauses but have the same polarity. Then these pairs trivially satisfy condition 3 of the **EDP** definition. Hence $\varphi \in \mathbf{EDP}_\Sigma(\sigma)$.
If $P \in \sigma$, then $\varphi \notin \mathbf{EDP}_\Sigma(\sigma)$ for any such $\sigma$. This is because $P$ is existential and condition 1 in Definition 2 allows only free or universal predicates with arity $\geq 2$ to be in $\sigma$.
As a second example, consider
$\psi = \exists x \forall z \exists v (P(v,z) \lor Q(z)) \land (P(x,v) \lor \neg Q(v))$
Here, $\Sigma = \{P, Q\}$, $\mathcal{F} = \mathcal{A} = \{\}$, $\mathsf{U} = \{Q\}$. Both $P$ and $Q$ have two instances in different clauses and for each of $P$ and $Q$, one of its instances is existential. But for $P$, both its instances are of the same polarity. For $Q$, we find that its instances even have opposite polarities, but $Q$ is unary. Then both $P$ and $Q$ satisfy condition 3 of Definition 2 trivially. Hence $\psi(x) \in \mathbf{EDP}_\Sigma(\{Q\})$ and $\psi(x) \notin \mathbf{EDP}_\Sigma(\Sigma)$.

**Lemma 10** *Let $\varphi$ be a $\mathbf{FO}_\Sigma$ sentence in PCNF in which*

1. *For every existential predicate $P$ of arity $\geq 2$ in $\Sigma$, either (a) all instances of $P$ in $\varphi$ have the same polarity or (b) all instances of $P$ appear in a single clause.*

2. *The equality predicate, if present, has free or universal arguments in all its instances.*

*Then there exists $\sigma \subseteq \Sigma$ such that $\varphi \in \mathbf{EDP}_\Sigma(\sigma)$.*

*Proof:* Let $\sigma \subseteq \Sigma$ be the (possibly empty) set of unary predicates alongwith the predicates of arity $\geq 2$ that are either free or universal in $\varphi$. Then it easily follows from Definition 2 that $\varphi \in \mathbf{EDP}_\Sigma(\sigma)$.

## 8.1 The $\mathbf{EDP}_\Sigma(\sigma)$ class as a syntactic fragment of the $\mathbf{EBS}_\Sigma(\sigma)$ class

**Theorem 4** *Let $\varphi$ be a $\mathbf{FO}_\Sigma$ sentence in $\mathbf{EDP}_\Sigma(\sigma)$. Let $k$ be the number of unary predicates in $\Sigma$. Then $\varphi$ satisfies all conditions in Definition 1 (of section 2) with $\mathcal{B} = |V(\varphi)| + |\overline{E_\mathsf{U}}| + 2^k$.*

Before proving this result, we see that this theorem immediately implies the following.

**Corollary 4** $\forall \sigma \subseteq \Sigma$, $\mathbf{EDP}_\Sigma(\sigma) \subseteq \mathbf{EBS}_\Sigma(\sigma)$.

Thus $\mathbf{EDP}_\Sigma(\sigma)$ is a syntactic fragment of $\mathbf{EBS}_\Sigma(\sigma)$. We shall explore the quesrion of whether $\mathbf{EDP}_\Sigma(\sigma)$ is a syntactic characterization of $\mathbf{EBS}_\Sigma(\sigma)$ in the forthcoming sections. We now present the proof of Theorem 4. The proof is a bit long so while the interested reader is urged to go through it, the proof if skipped would not affect the flow.

### 8.1.1 Proving Theorem 4

We prove a stronger version of Theorem 4 namely for formulae with free variables. The results naturally follows for sentences too.
For the case of free variables, the **EBS** definition is slightly widened to also assert that if $M \models \varphi(\mathbf{a})$ where $\mathbf{a}$ is a vector of elements from $M$, then $M_1$ must contain $\mathbf{a}$. The other conditions remain exactly the same.

Our proof is motivated by the proof of finite model size of formulae in the Löwenheim class and the Bernays-Schönfinkel-Ramsey class, as given in [2].
Consider a $\Sigma$−structure $M$ with universe $\mathcal{U}_M$ and $\mathbf{a} \in \mathcal{U}_M^{|\mathbf{x}|}$ such that $M \models \varphi(\mathbf{a})$ and $|\mathcal{U}_M| > m + |V(\varphi(\mathbf{x}))| + |\overline{E_\mathsf{U}}| + 2^k$.
Now we have that $|\mathcal{C}| = m$ and $|\mathsf{U}| = k$. Let $A' \subseteq \mathcal{U}_M$ be the set of interpretations of the constants of $\mathcal{C}$ in the model $M$. Let $A'' \subseteq \mathcal{U}_M$ be the set of values assigned to variables in $V(\varphi(\mathbf{x}))$ in showing that $M \models \varphi(\mathbf{a})$ (The existential variables in $EV(\varphi(\mathbf{x}))$ being the leftmost, the values assigned to them in $M$ are independent of the values of all the universal variables). Thus, $A''$ contains all values in $\mathbf{a}$ and also values assigned to variables bound to the leftmost existential quantifiers in $\varphi(\mathbf{x})$. Clearly, $|A'| \leq m$ and $|A''| \leq |V(\varphi(\mathbf{x}))|$. Since as per our terminology, all the arguments belonging to $V(\varphi(\mathbf{x})) \cup \mathcal{C}$ are called *free* arguments, we denote by *Free* the set $V(\varphi(\mathbf{x})) \cup \mathcal{C}$ and by **Val**(*Free*) the set of values in $M$, for the elements of *Free* i.e. **Val**(*Free*) $= A' \cup A''$.

Without loss of generality, let $\mathsf{U} = \{Q_1, \ldots, Q_k\}$. For every element $e$ in the universe $\mathcal{U}_M$, we associate a binary vector or *colour* $C(e) \in \{0, 1\}^k$, where the $i^{th}$ component of $C(e)$ is 1 iff $M \models Q_i(e)$. For every $c \in \{0, 1\}^k$, let $\mathcal{U}_M^c$ denote the subset of elements of $\mathcal{U}_M$ that have the colour $c$. Consider the set $\mathcal{V}_M^c = \mathcal{U}_M^c \setminus \mathbf{Val}(Free)$. We now choose a subset $A^c$ of $\mathcal{V}_M^c$ as follows: if $\mathcal{V}_M^c = \emptyset$, then $A^c = \mathcal{V}_M^c = \emptyset$; otherwise, $A^c$ is a singleton set formed by selecting any element $a^c$ from $\mathcal{V}_M^c$ (i.e. $A^c = \{a^c\}$).
Let $\overline{E_\mathsf{U}} = EV(\varphi(\mathbf{x})) \setminus E_\mathsf{U}$ denote the complement of $E_\mathsf{U}$ within $EV(\varphi(\mathbf{x}))$. $\overline{E_\mathsf{U}}$ denotes the set of all variables of $EV(\varphi(\mathbf{x}))$ that do not appear in $\varphi(\mathbf{x})$ as an argument of any instance of any unary predicate. Then we finally choose a subset of $\mathcal{U}_M$ of size $|\overline{E_\mathsf{U}}|$ which we denote as **Val**($\overline{E_\mathsf{U}}$) such that (i) $\mathbf{Val}(\overline{E_\mathsf{U}}) \cap \mathbf{Val}(\textit{Free}) = \emptyset$ and (ii) $\mathbf{Val}(\overline{E_\mathsf{U}}) \cap A^c = \emptyset$ for all colours $c \in \{0,1\}^k$.
Note that

$$|\mathbf{Val}(\textit{Free})| + |\mathbf{Val}(\overline{E_\mathsf{U}})| + \Sigma_{c \in \{0,1\}^k}|A^c|$$
$$\leq |A'| + |A''| + |\overline{E_\mathsf{U}}| + \Sigma_{c \in \{0,1\}^k} 1$$
$$\leq m + |V(\varphi(\mathbf{x}))| + |\overline{E_\mathsf{U}}| + 2^k$$

so that we can indeed choose the subset $\mathbf{Val}(\overline{E_\mathsf{U}})$ of $\mathcal{U}_M$ of the mentioned size. We denote the elements of $\mathbf{Val}(\overline{E_\mathsf{U}})$ as $a_i$ where $i$ is such that $v_i \in \overline{E_\mathsf{U}}$ (thus $\mathbf{Val}(\overline{E_\mathsf{U}}) = \{a_i | v_i \in \overline{E_\mathsf{U}}\}$). Note that the sets $\mathbf{Val}(\textit{Free})$, $\mathbf{Val}(\overline{E_\mathsf{U}})$ and $A^c$ are all disjoint with each other for each $c \in \{0,1\}^k$. Further $A^{c_1}$ and $A^{c_2}$ are also disjoint with each other for all $c_1, c_2 \in \{0,1\}^k$ where $c_1 \neq c_2$.

We now consider the sub-structure $M_1$ of $M$ generated by $\mathbf{Val}(\textit{Free}) \cup \mathbf{Val}(\overline{E_\mathsf{U}}) \cup \bigcup_{c \in \{0,1\}^k} A^c$. We show that this sub-structure $M_1$ satisfies all of the **EBS** conditions with $\mathcal{B} = m + |V(\varphi(\mathbf{x}))| + |\overline{E_\mathsf{U}}| + 2^k$.
Since $\varphi(\mathbf{x})$ is in prenex normal form, we assume without loss of generality that

$$\varphi(\mathbf{x}) = \exists \mathbf{v}_{\textit{leftmost}} \forall \mathbf{z}_0 \exists v_1 \forall \mathbf{z}_1 \exists v_2 \forall \mathbf{z}_2 \ldots \exists v_r \forall \mathbf{z}_r$$
$$\psi(\mathbf{x}, \mathbf{v}_{\textit{leftmost}}, \mathbf{z}_0, v_1, \mathbf{z}_1, \ldots v_r, \mathbf{z}_r)$$

where $\psi$ is a quantifier-free matrix in CNF, $\mathbf{v}_{\textit{leftmost}}$ denotes the (possibly empty) set of variables bound to the leftmost quantifiers (all of these variables are thus in $V(\varphi(\mathbf{x}))$), $r$ denotes $|EV(\varphi(\mathbf{x}))|$ and $\mathbf{z}_i$ denotes the (possibly empty) vector of universally quantified variables that appear immediately to the right of the existentially quantified variable $v_i$ in the quantifier prefix of $\varphi(\mathbf{x})$. Let $\mathbf{z} = (\mathbf{z}_0, \mathbf{z}_1, \ldots, \mathbf{z}_r)$ denote the tuple of all the $AV(\varphi(\mathbf{x}))$ variables and $\mathbf{v} = (v_1, v_2, \ldots, v_r)$ denote the tuple of all the $EV(\varphi(\mathbf{x}))$ variables. For every extension $M_2$ of $M_1$ within $M$, we now describe how to construct a $\Sigma$-structure $M_2'$ such that $\mathcal{U}_{M_2} = \mathcal{U}_{M_2'}$, $M_2' \models \varphi(\mathbf{a})$ and $M_2'|_\sigma = M_2|_\sigma$.

If $M_2 \models \varphi(\mathbf{a})$, we simply choose $M_2' = M_2$. Clearly, this choice satisfies the **EBS** conditions. If $M_2 \not\models \varphi(\mathbf{a})$, we must redefine the interpretations of (some) predicates in $\Sigma \setminus \sigma$ such that the resulting structure $M_2'$ is a model of $\varphi(\mathbf{a})$. We note that this $M_2'$ has the same universe namely $\mathcal{U}_{M_2}$ as $M$ i.e. $\mathcal{U}_{M_2'} = \mathcal{U}_{M_2}$. Therefore henceforth we will refer to the universe of $M_2'$ as $\mathcal{U}_{M_2}$.

The way we proceed is we first create a *quasi* $\Sigma$-structure $M_3$ which has the universe $\mathcal{U}_{M_2}$ and such that $\varphi(\mathbf{a})$ is True in it. It is *quasi* in the sense that it gives *atleast* one truth value to each predicate of $\Sigma$ for each valuation of its arguments. In other words, it could give both truth values to a predicate for certain valuations of its arguments. The desired $M_2'$ would then be extracted out from this quasi $\Sigma$-structure $M_3$.

In the following, we overload the $\models$ symbol so that we denote $\varphi(\mathbf{a})$ being True in $M_3$ as $M_3 \models \varphi(\mathbf{a})$.
Now, in order to get $M_3$, we observe that for every assignment of values $\mathbf{Z}$ from $\mathcal{U}_{M_2}$ to the universally quantified variables $\mathbf{z}$ in $\varphi(\mathbf{a})$, we must choose a value from $\mathcal{U}_{M_2}$ for each of the constants in $\mathcal{C}$ and each of the existentially quantified variables, namely the variables in $\mathbf{v}_{\textit{leftmost}}$ and those in $\mathbf{v}$, such that $\psi(\mathbf{x}, \mathbf{v}_{\textit{leftmost}}, \mathbf{z}_0, v_1, \mathbf{z}_1, \ldots v_r, \mathbf{z}_r)$ instantiated with these values evaluates to True in $M_3$. Suppose we are indeed able to choose these values from $\mathcal{U}_{M_2}$. Upon substitution of these values in $\psi$, call the resulting fully instantiated matrix as $\psi_{M_3}[\mathbf{Z}]$ (Note the square brackets in the notation which are used to mean that $\mathbf{Z}$ is considered as an 'input' to give as 'output', a fully instantiated matrix denoted as $\psi_{M_3}[\mathbf{Z}]$. It should *not* be looked at as meaning that $\psi_{M_3}$ is a formula whose free variables are being instantiated with $\mathbf{Z}$ - in that case we would use round brackets around $\mathbf{Z}$).
Before proceeding ahead, we observe that there are two tasks to be done. The first is, given a valuation $\mathbf{Z}$ of the universal variables, choosing the values for the constants and all the existential variables to substitute in $\psi$ to get $\psi_{M_3}[\mathbf{Z}]$. The second is to define the (quasi) structure $M_3$ over which $\psi_{M_3}[\mathbf{Z}]$ evaluates to True for each valuation $\mathbf{Z}$ from $\mathcal{U}_{M_2}$ of the $\mathbf{z}$ variables so that as a result $M_3 \models \varphi(\mathbf{a})$.
For the first task, we choose the values as follows:

1. For each constant in $\mathcal{C}$, we interpret it in $M_3$ with the same value with which it is interpreted in $M$. Likewise for each variable of $V(\varphi((\mathbf{x}))$, we substitute for it the same value which is assigned to it in $M$ (This is possible since $\mathbf{Val}(\textit{Free}) \subseteq \mathcal{U}_{M_2}$). These values of course are independent of the instantiations $\mathbf{Z}$ for the $\mathbf{z}$ variables. This takes care of all elements of *Free*.

2. For each variable $v_i \in \overline{E_\mathsf{U}}$, we assign it the value $a_i \in \mathbf{Val}(\overline{E_\mathsf{U}}) \subseteq \mathcal{U}_{M_2}$. These values too are independent of the instantiations of the $\mathbf{z}$ variables.

3. Given a valuation $\mathbf{Z}$ of the $\mathbf{z}$ variables, for each variable $v_i \in E_\mathsf{U}$, we do the following. We turn to the model $M$ and look at the assignment made to the $\mathbf{v}$ variables in $\psi$ for this valuation, namely $\mathbf{Z}$, of the $\mathbf{z}$ variables. Since $M \models \varphi(\mathbf{a})$, we know that there exists a choice of values from $\mathcal{U}_M$ for each existentially quantified variable and the constants such that when these values are substituted in $\psi(\mathbf{x}, \mathbf{v}_{\textit{leftmost}}, \mathbf{z}_0, v_1, \mathbf{z}_1, \ldots v_r, \mathbf{z}_r)$, we get a fully instantiated matrix - which we denote as $\psi_M[\mathbf{Z}]$ - which evaluates to True in $M$. Let $d$ be the value substituted for $v_i$ from $\mathcal{U}_M$ in making $\psi_M[\mathbf{Z}]$ True in $M$. Let $c$ be the colour of $d$ in $M$.

    (a) If $d$ belongs to $\mathbf{Val}(\textit{Free})$, then we choose the

same value $d$ to substitute for $v_i$ from $\mathcal{U}_{M_2}$ (This is possible because **Val**(*Free*) $\subseteq \mathcal{U}_{M_2}$).

(b) If $d$ is not in **Val**(*Free*), then we choose the value $a^c \in A^c \subseteq \mathcal{U}_{M_2}$ to substitute for $v_i$.

Using the above procedure (call it *Selector*), for each valuation $\mathbf{Z}$ of the $\mathbf{z}$ variables, we determine the values to assign to each of the existential variables and constants and substituting all these values in $\psi$, we get the fully instantitated matrix $\psi_{M_3}[\mathbf{Z}]$.
We now take up the second task namely defining $M_3$ such that $M_3 \models \varphi(\mathbf{a})$.

Before proceeding ahead, we introduce a little terminology to make further explanations easier. If $I$ denotes the instance of a predicate $P$ in $\psi$, we denote by $I_M[\mathbf{Z}]$, the instance $I$ as it appears in $\psi_M[\mathbf{Z}]$ (Thus $I_M[\mathbf{Z}]$ would look like $P(\mathbf{d})$ where the values in $\mathbf{d}$ come from $\mathcal{U}_M$). Likewise we denote by $I_{M_3}[\mathbf{Z}]$, the instance $I$ as it appears in $\psi_{M_3}[\mathbf{Z}]$. Then for each valuation $\mathbf{Z}$ of the $\mathbf{z}$ variables, for each predicate $P \in \Sigma \cup \{=\}$ and for each instance $I$ of $P$ in $\psi$, we do the following:

1. If $P$ is nullary (i.e. constant), then we have already stated above how $P$ must be interpreted in $M_3$.

2. If $P \in \mathsf{U}$, then $I_{M_3}[\mathbf{Z}]$ is assigned the same truth value in $M_3$ as $I_M[\mathbf{Z}]$ has in $M$. In other words $M_3 \models I_{M_3}[\mathbf{Z}]$ iff $M \models I_M[\mathbf{Z}]$.

3. If $P$ is atleast 2-ary, we have the following:

    (a) If $P$ is a free or universal predicate, then $I_{M_3}[\mathbf{Z}]$ is assigned a truth value such that $M_3 \models I_{M_3}[\mathbf{Z}]$ iff $M \models I_M[\mathbf{Z}]$.

    (b) If $P$ is an existential predicate, then $I_{M_3}[\mathbf{Z}]$ is assigned a truth value such that
    - $M_3 \models I_{M_3}[\mathbf{Z}]$ if $I$ is $+ve$ in $\psi$ and
    - $M_3 \models \neg I_{M_3}[\mathbf{Z}]$ if $I$ is $-ve$ in $\psi$.

By the above, for each valuation $\mathbf{Z}$ of $\mathbf{z}$, since either we have given an instance in $\psi_{M_3}[\mathbf{Z}]$ the same truth value in $M_3$ as the corresponding instance in $\psi_M[\mathbf{Z}]$ has in $M$ or we have made the literal corresponding to the instance True in $M_3$, we can see that $\psi_{M_3}[\mathbf{Z}]$ is true in $M_3$ since $\psi_M[\mathbf{Z}]$ is true in $M$.
We note now that by the above procedure, we would have *interpreted* in $M_3$ (i.e. given truth values in $M_3$ to) all the predicates in $\Sigma$ for those values of their arguments which were seen in the '$\psi_{M_3}[\mathbf{Z}]$'s that were obtained when enlisting all the valuations $\mathbf{Z}$ of the universally quantified variables. Lets call this whole procedure as procedure $\mathcal{P}_1$.
In general, $\mathcal{P}_1$ might not exhaust all possible argument values for all the predicates and hence the interpretation of the predicates in $M_3$ would be partial and incomplete. In order to give the complete interpretation in $M_3$ for each predicate in $\Sigma$, we need to consider for each predicate, those values for its arguments that were not covered by $\mathcal{P}_1$. As for the interpretations of the predicates for these values (i.e. which were not covered by $\mathcal{P}_1$), it doesnt matter in which way they are interpreted because these values do not in any way contribute towards making $\psi_{M_3}[\mathbf{Z}]$ True in $M_3$ for any valuation $\mathbf{Z}$ from $\mathcal{U}_{M_2}$ of the $\mathbf{z}$ variables (and hence in showing $M_3 \models \varphi(\mathbf{a})$). Hence, in particular, they could be interpreted as they are in $M$. Lets call this procedure of assignment of truth values to the predicates of $\Sigma$ for their remaining argument values as procedure $\mathcal{P}_2$. Clearly procedures $\mathcal{P}_1$ and $\mathcal{P}_2$ completely define in $M_3$, the interpretations of all predicates for all values of their arguments. Further they define interpretations of the predicates for disjoint sets of values of their arguments.

Having defined $M_3$, we now see that it is possible that above procedures $\mathcal{P}_1$ and $\mathcal{P}_2$ together give interpretations to the predicates of $\Sigma$ in $M_3$ that cause a *conflict* i.e. a predicate being assigned both True and False values in $M_3$ for the same values of its arguments. Conflicts thus make $M_3$ a quasi $\Sigma-$structure (and not a $\Sigma-$structure). In what follows we show that either conflicts are *prevented* by the **EDP** conditions or conflicts that are not prevented can be *cured* (resolved) without affecting the truth of $\varphi(\mathbf{a})$ in the 'cured' $M_3$. In other words, we will be able to construct the desired model $M_2'$ by simply curing all the conflicts in $M_3$ (i.e. by extracting the appropriate truth value in case of a conflict). Then $M_2'$ would be a valid $\Sigma-$structure giving unique truth values to each predicate for each valuation of its arguments. Further $M_2'$ would be a model for $\varphi(\mathbf{a})$ i.e. $M_2' \models \varphi(\mathbf{a})$.
Before going ahead, we note that $M$ is clearly free of any conflicts.

**Handling conflicts**

Suppose there was a conflict i.e. a predicate $P$ got assigned both True and False values in $M_3$ for the same vector of values (say $\mathbf{d}$) as its argument. Then we have the following cases:
1) Suppose the two different truth values to $P(\mathbf{d})$ are both assigned by $\mathcal{P}_2$. But $\mathcal{P}_2$ assigns the same truth values as in $M$. This shows that there must be a conflict in $M$ itself which we know is not the case.
2) Without loss of generality, suppose $P(\mathbf{d})$ is assigned True by $\mathcal{P}_1$ and False by $\mathcal{P}_2$. Then as seen above, since $\mathcal{P}_1$ and $\mathcal{P}_2$ give truth values for disjoint sets of values of arguments, $P(\mathbf{d})$ would be assigned a truth value either only by $\mathcal{P}_1$ or only by $\mathcal{P}_2$ contradicting the assumption above.
3) $P(\mathbf{d})$ is assigned both the truth values by $\mathcal{P}_1$. We abbreviate the truth value True as $t$ and False as $f$. Having seen how the procedure $\mathcal{P}_1$ works, we conclude that for each truth value $\alpha \in \{t, f\}$, $P(\mathbf{d})$ must then have appeared in $\psi_{M_3}[\mathbf{Z}_\alpha]$ for some valuation $\mathbf{Z}_\alpha$ of the $\mathbf{z}$ variables where it was assigned that truth value (namely $\alpha$). Let for truth value $\alpha$, the instance $I^\alpha$ of $P$ in $\psi$ be instantiated in $\psi_{M_3}[\mathbf{Z}_\alpha]$ to get $P(\mathbf{d})$. In particular, the instance $I^t$ of $P$ in $\psi$ gets instantiated in $\psi_{M_3}[\mathbf{Z}_t]$ to get $P(\mathbf{d})$ and the instance $I^f$ of $P$ in $\psi$ gets instantiated in $\psi_{M_3}[\mathbf{Z}_f]$ to get $P(\mathbf{d})$.

Using our notation introduced earlier, the instances mentioned above would be denoted as $I^t_{M_3}[\mathbf{Z}_t]$ and $I^f_{M_3}[\mathbf{Z}_f]$ respectively ($I^t_{M_3}[\mathbf{Z}_t] = I^f_{M_3}[\mathbf{Z}_f] = P(\mathbf{d})$). Further the former instance is assigned truth value as True and the latter is assigned False in $M_3$.

Before proceeding ahead, we make the following observation.

**Lemma 11** *Consider an instance $I$ in $\psi$ of a predicate $P$. Let $\mathbf{Z}$ be an instantiation from $\mathcal{U}_{M_2}$ of all the universal variables $\mathbf{z}$. Then the following are true:*

1. *For each $i$ ranging from 1 to the arity of $P$, if either (i) the $i^{th}$ argument of $I$ in $\psi$ is universal or (ii) the $i^{th}$ argument of $I_{M_3}[\mathbf{Z}]$ is from **Val**(Free), then the $i^{th}$ argument of $I_M[\mathbf{Z}]$ is the same as the $i^{th}$ argument of $I_{M_3}[\mathbf{Z}]$.*

2. *If $I$ is a free or a universal instance in $\psi$, then $I_M[\mathbf{Z}]$ and $I_{M_3}[\mathbf{Z}]$ have the same vector of values appearing as their arguments.*

*Proof:*

1. (i) If the $i^{th}$ argument is universal, then since $\psi_{M_3}[\mathbf{Z}]$ and $\psi_M[\mathbf{Z}]$ have the same valuation of all the universal variables, namely $\mathbf{Z}$, then in particular for the $i^{th}$ argument of $I$, the same value is substituted for it in both $\psi_{M_3}[\mathbf{Z}]$ and $\psi_M[\mathbf{Z}]$.

   (ii) If the $i^{th}$ argument of $I_{M_3}[\mathbf{Z}]$ is from **Val**(Free), then it means that the $i^{th}$ argument of $I$ is either (a) free or (b) universal or (c) a variable in $E_\mathsf{U}$. For (a) and (c), we see directly from cases (1) and (3a) of *Selector* that the $i^{th}$ argument of $I_M[\mathbf{Z}]$ is the same as the $i^{th}$ argument of $I_{M_3}[\mathbf{Z}]$. For (b), the reasoning is already presented above.

2. If $I$ is a free or a universal instance, each argument of it is either free or universal. Then by applying the reasoning presented in the preceding point to all arguments $i$ ranging from 1 to the arity of $P$, we can see that $I_M[\mathbf{Z}]$ and $I_{M_3}[\mathbf{Z}]$ will have the same vector of values appearing as their arguments.

∎

Coming back to checking for conflicts, we have the following possibilities:

1. $P$ is unary. Then $\mathbf{d} = d$ (say).

   (a) If $I^t$ and $I^f$ are free or universal instances, then from Lemma 11(2), for each truth value $\alpha \in \{t, f\}$, $I^\alpha_M[\mathbf{Z}_\alpha]$ and $I^\alpha_{M_3}[\mathbf{Z}_\alpha]$ have the same value as their argument (namely $d$). Then $I^t_M[\mathbf{Z}_t] = I^f_M[\mathbf{Z}_f] = P(d)$.

   (b) If $d \in$ **Val**(Free), then from Lemma 11(1), we again see similarly that $I^t_M[\mathbf{Z}_t] = I^f_M[\mathbf{Z}_f] = P(d)$.

   In either of the above cases, from case (2) of $\mathcal{P}_1$, $I^\alpha_M[\mathbf{Z}_\alpha]$ and $I^\alpha_{M_3}[\mathbf{Z}_\alpha]$ have the same truth value (namely $\alpha$) in $M$ and $M_3$ respectively. But then this shows that $P(d)$ is assigned both truth values in $M$ thus showing a conflict in $M$ itself which we know does not exist.

   Without loss of generality, suppose $I^t$ is an existential instance (i.e. of the form $P(v)$ where $v \in EV(\varphi(\mathbf{x}))$). Note that the argument $v$ cannot belong to $\overline{E_\mathsf{U}}$. The case when $v$ is assigned the value $d \in$ **Val**(Free) is already considered above. Then from case 3(b) of *Selector*, the only remaining case is when $d = a^c$ for some colour $c$. Since $a^c$ appears as the argument of $I^t_{M_3}[\mathbf{Z}_t]$, from case (3b) of *Selector*, it means that the colour of $a^c$ in $M$ is $c$. If $I^t_M[\mathbf{Z}_t] = P(d_1)$, then the colour of $d_1$ is also $c$. Further since by case(2) of $\mathcal{P}_1$, $M \models I^t_M[\mathbf{Z}_t]$ iff $M_3 \models I^t_{M_3}[\mathbf{Z}_t]$, we have that $P(d_1)$ is True in $M$. Then since both $d_1$ and $a^c$ have the same colour in $M$, we have that $P(a^c)$ is also True in $M$.

   Since the argument of $I^f_{M_3}[\mathbf{Z}_f]$ is also $a^c$, it means that the argument of $I^f$ is either (i) universal or (ii) existential (in particular, belonging to $E_\mathsf{U}$. In case of (i), by Lemma 11(2), $I^f_M[\mathbf{Z}_f] = I^f_{M_3}[\mathbf{Z}_f] = P(a^c)$. By case (2) of $\mathcal{P}_1$, $M \models I^f_M[\mathbf{Z}_f]$ iff $M_3 \models I^f_{M_3}[\mathbf{Z}_f]$, so that $P(a^c)$ is False in $M$. But this is a contradiction.

   In case of (ii), suppose $I^f_M[\mathbf{Z}_f] = P(d_2)$. Since $I^f_{M_3}[\mathbf{Z}_f] = P(a^c)$, by case (3b) of *Selector*, $d_2$ has the colour $c$. Then $P(d_2)$ is True in $M$. But by case (2) of $\mathcal{P}_1$, $M \models I^f_M[\mathbf{Z}_f]$ iff $M_3 \models I^f_{M_3}[\mathbf{Z}_f]$ so that $P(d_2)$ is False in $M$. Now since $d_2$ and $a^c$ have the same colour in $M$, we must have that $P(a^c)$ is False in $M$. This is again a contradiction.

   Thus there are no conflicts in $M_3$ as far as the unary predicates go.

2. $P$ is of arity $\geq 2$ and is a free or universal predicate.

   Then each of $I^t$ and $I^f$ is either universal or free in $\varphi(\mathbf{x})$. Then from Lemma 11(2), for each truth value $\alpha \in \{t, f\}$, $I^\alpha_M[\mathbf{Z}_\alpha]$ and $I^\alpha_{M_3}[\mathbf{Z}_\alpha]$ have the same vector of values as their arguments (namely $\mathbf{d}$). Further by case (3a) of $\mathcal{P}_1$, both have the same truth value (namely $\alpha$) in $M$ and $M_3$ respectively. But then this shows that $I^t_M[\mathbf{Z}_t] = I^f_M[\mathbf{Z}_f] = P(\mathbf{d})$ and that $P(\mathbf{d})$ is assigned both truth values in $M$ thus showing a conflict in $M$ itself which we know does not exist.

Thus there are no conflicts in $M_3$ for the free and universal predicates too.

3. $P$ is of arity $\geq 2$ and is an existential predicate.

   If $I^t$ and $I^f$ are of the same polarity in $\varphi(\mathbf{x})$, then from case (3b) of $\mathcal{P}_1$, we see that in $M_3$, both $I^t_{M_3}[\mathbf{Z}_t]$ and $I^f_{M_3}[\mathbf{Z}_f]$ have the same truth value - True if their polarity is $+ve$ and False if their polarity is $-ve$. Then there is no conflict in the first place.

   If $I^t$ and $I^f$ are of opposite polarity, then suppose that they were in different clauses. From the last condition of the **EDP** definition, there is a variable $v_l \in \overline{E_\mathsf{U}}$ such that $I^t$ and $I^f$ would be existentially distinguishable by it. Suppose it appears as the $n^{th}$ argument of $I^f$. Then the $n^{th}$ argument of $I^t$ would be either an variable $v \in$ *Free* or a variable $v_p \in E_\mathsf{U}$ or a variable $v_r \in \overline{E_\mathsf{U}}$ where $r \neq l$. Then *Selector* would assign the $n^{th}$ argument of $I^f_{M_3}[\mathbf{Z}_f]$ as $a_l \in \mathbf{Val}(\overline{E_\mathsf{U}})$ while the $n^{th}$ argument of $I^t_{M_3}[\mathbf{Z}_t]$ would be respectively assigned a value either from (i) $\mathbf{Val}(Free)$ or from (ii) $A^c \neq \emptyset$ for some colour $c$ or (iii) $a_r \in \mathbf{Val}(\overline{E_\mathsf{U}})$. The values in case of (i) and (ii) must be different from $a_l$ since $\mathbf{Val}(\overline{E_\mathsf{U}})$ is disjoint with both $\mathbf{Val}(Free)$ and $A^c$. In case of (iii), since $r \neq l$, $a_r \neq a_l$. Thus in all cases, the $n^{th}$ argument of $I^f_{M_3}[\mathbf{Z}_f]$ would be different from that of $I^t_{M_3}[\mathbf{Z}_t]$. Then the conflict couldn't have happened at all. A similar argument can be given for the case when $v_l$ appears as the $n^{th}$ argument of $I^t$.

   (We thus note how the last condition in the **EDP** definition *prevents* the occurence of a conflict.)

   Finally consider the case that $I^t$ and $I^f$ are of opposite polarity and are in the same clause say $C$ of $\psi$. Then since $M_3 \models I^t_{M_3}[\mathbf{Z}_t]$ and $M_3 \models \neg I^f_{M_3}[\mathbf{Z}_f]$, by case (3b) of $\mathcal{P}_1$, we have that $I^t$ is $+ve$ while $I^f$ is $-ve$ in $\varphi(\mathbf{x})$. Then for the valuation $\mathbf{Z}_f$ of the $\mathbf{z}$ variables, $\mathcal{P}_1$ would assign $I^f_{M_3}[\mathbf{Z}_f]$, the value True in $M_3$. Let $C[\mathbf{Z}_f]$ be the clause $C$ as it appears in $\psi_{M_3}[\mathbf{Z}_f]$. Then $C[\mathbf{Z}_f]$ which contains $I^t_{M_3}[\mathbf{Z}_f]$ and $I^f_{M_3}[\mathbf{Z}_f]$, is not dependent on the literal corresponding to $I^f_{M_3}[\mathbf{Z}_f]$ for $C[\mathbf{Z}_f]$ to become True in $M_3$ as $I^t_{M_3}[\mathbf{Z}_f]$ is already True in $C[\mathbf{Z}_f]$. Then we *cure* the conflict on the value of $P(\mathbf{d})$ ($= I^t_{M_3}[\mathbf{Z}_t] = I^f_{M_3}[\mathbf{Z}_f]$) by fixing the value of $P(\mathbf{d})$ in $M_3$ to be True.

   Now since $M_3 \models \varphi(\mathbf{a})$, we have that for every clause $C_1$ of $\psi$, $C_1[\mathbf{Z}]$ is true in $M_3$ for each valuation $\mathbf{Z}$ of the $\mathbf{z}$ variables ($C_1[\mathbf{Z}]$ is the clause $C_1$ as it appears in $\psi_{M_3}[\mathbf{Z}]$). We note that while the fixing mentioned above doesnt affect the truth of clause $C[\mathbf{Z}_f]$, we now investigate whether this makes some other clause False i.e. does there exist a valuation $\mathbf{Z}^*$ of $\mathbf{z}$ and a clause $C^*$ of $\psi$ such that in $\psi_{M_3}[\mathbf{Z}^*]$, the clause $C^*[\mathbf{Z}^*]$ (which is the clause $C^*$ as it appears in $\psi_{M_3}[\mathbf{Z}^*]$) became False due to the fixing. Suppose there is such a $\mathbf{Z}^*$. Since the fixing caused the change of truth value of $C^*[\mathbf{Z}^*]$ from True to False, it must be the case that there is an instance $I^*$ of $P$ appearing in $C^*[\mathbf{Z}^*]$ such that $I^*_{M_3}[\mathbf{Z}^*] = P(\mathbf{d})$ and such that $I^*$ is $-ve$ in $\psi$. Further $I^*_{M_3}[\mathbf{Z}^*]$ was earlier (i.e. before the fixing) assigned value False (making its corresponding literal True) and every other literal in $C^*[\mathbf{Z}^*]$ is assigned value False.

We firstly note that $I^* \neq I^t$ since they are of opposite polarities. If $C^* \neq C$, then $I^t$ and $I^*$ are two instances of $P$ of opposite polarity appearing in different clauses, which had earlier (i.e. before the fixing) conflicted. But in the analysis shown above we know that this cannot happen. Then $C^* = C$. But then by case (3b) of $\mathcal{P}_1$, $I^t_{M_3}[\mathbf{Z}^*]$ would be assigned True in $M_3$. But this contradicts the conclusion made above that the literal corresponding to $I^*_{M_3}[\mathbf{Z}^*]$ was the only literal in $C^*[\mathbf{Z}^*]$ that made the latter True.

Thus, by fixing the value of $P(\mathbf{d})$ to True, the conflict is eliminated and further for every clause $C$ in $\psi$ and for every valuation $\mathbf{Z}$ of $\mathbf{z}$, $C[\mathbf{Z}]$ continues to remain True so that $\varphi(\mathbf{a})$ continues to be True in the new $M_3$ which is cured of one conflict.

Since $M_3$ is of finite size, there are only a finite number of conflicts and the curing process above eliminates conflicts one at a time while still preserving the truth of $\varphi(\mathbf{a})$ in the new $M_3$ obtained after the curing. Then this process terminates, giving a *fully cured $M_3$* which we call $M'_2$ such that $M'_2$ has no conflicts in it at all (and hence is a valid $\Sigma-$structure) and such that $M'_2 \models \varphi(\mathbf{a})$. We thus note that $M'_2$ has been obtained by simply resolving all the conflict cases in $M_3$ and retaining the truth values of all predicates for all their arguments for which there is no conflict.

**Completing the proof**

With a last remaining argument now, we show that with this $M'_2$, satisfies the last condition of the **EBS** definition.
For the proof to complete we need to show that $M'_2|_\sigma = M_2|_\sigma$ (recall that $M_2$ is the substructure of $M$ generated by $\mathcal{U}_{M_2}$). Since $\sigma$ can contain only nullary, unary, free and existential predicates, we will consider each of these separately.

1. Nullary predicates: From case (1) of *Selector*, all the elements of $\mathcal{C}$ are interpreted in $M_3$, and hence in $M'_2$, with the same values as they are interpreted in $M$. Thus $M'_2|_\mathcal{C} = M_2|_\mathcal{C}$.

2. Free and universal predicates: For a free or universal predicate $P$ of arity $\geq 2$, using Lemma 11(2), $\mathcal{P}_2$, case (3a) of $\mathcal{P}_1$ and the conclusion made above that there are no conflicts on free and universal predicates, we can conclude that $M_3|_{\{P\}} = M_2|_{\{P\}}$. Since in $M'_2$ is obtained from $M_3$ simply by fixing the conflict cases and retaining the truth values of all predicates for all their arguments for which there is no conflict, we conclude that $M'_2|_{\{P\}} = M_3|_{\{P\}} = M_2|_{\{P\}}$.

3. Unary predicates: For $P \in \mathsf{U}$, for the following cases its is easy to see that $M \models P(e)$ iff $M_3 \models P(e)$: (i) $P(e)$ is assigned the truth value by $\mathcal{P}_2$ (ii) $P(e) = I_{M_3}[\mathbf{Z}]$ for some valuation $\mathbf{Z}$ of the universal variables and some instance $I$ of $P$ which is free or universal.

If $P(e) = I_{M_3}[\mathbf{Z}]$ where $I$ is existential, then for $e \in \mathbf{Val}(\mathit{Free})$, it is easy to see from Lemma 11(1) and case (2) of $\mathcal{P}_1$, that $M \models P(e)$ iff $M_3 \models P(e)$. Else $e = a^c$ for some colour $c$. Then if $I_M[\mathbf{Z}] = P(d_1)$, we have already seen above that $d_1$ and $a^c$ have the same colour $c$ in $M$. Putting this together with the fact from case (2) of $\mathcal{P}_1$ that $M \models I_M[\mathbf{Z}]$ iff $M_3 \models I_{M_3}[\mathbf{Z}]$, we have that $P(e)$ has the same truth value in $M$ and $M_3$. Thus in all cases $M \models P(e)$ iff $M_3 \models P(e)$. Since there are no conflicts on unary predicates, $M_3|_{\{P\}} = M_2|_{\{P\}}$. Further since $M'_2$ is obtained from $M_3$ simply by fixing the conflict cases and retaining the truth values of all predicates for all their arguments for which there is no conflict, we conclude that $M'_2|_{\{P\}} = M_3|_{\{P\}} = M_2|_{\{P\}}$.

We conclude from the above that $M'_2|_\sigma = M_2|_\sigma$ and indeed, that completes the proof! ∎

## 8.2 Some observations about the EDP class

We can see that both $\mathbf{BSR}_\Sigma$ and $\mathbf{L}_\Sigma$ without equality (i.e. just the Löwenheim class) lie wholly within $\mathbf{EDP}_\Sigma(\Sigma)$. In case of $\varphi \in \mathbf{BSR}_\Sigma$, $EV(\varphi) = \emptyset$ and hence it contains no existential predicates at all. Hence all the conditions of the $\mathbf{EDP}_\Sigma(\Sigma)$ definition are trivially satisfied. In case of a sentence $\varphi$ in $\mathbf{L}_\Sigma$ without equality, it contains no predicates of arity $\geq 2$ and no equality too. Hence all conditions of the $\mathbf{EDP}_\Sigma(\Sigma)$ definition are again trivially satisfied. Thus $\mathbf{BSR}_\Sigma \subseteq \mathbf{EDP}_\Sigma(\Sigma)$ and $\mathbf{L}_\Sigma$ without equality $\subseteq \mathbf{EDP}_\Sigma(\Sigma)$.

Let for sets $A, B$ of sentences, $A \sqsubseteq B$ denote that $A$ is semantically contained inside $B$ i.e. every sentence in $A$ is semantically equivalent to some sentence in $B$. We then have the following.

**Theorem 5** *1. If $\sigma_1 \subseteq \sigma_2 \subseteq \Sigma$, then $\mathbf{EDP}_\Sigma(\sigma_2) \subseteq \mathbf{EDP}_\Sigma(\sigma_1)$.*

*2. $\mathbf{EDP}_\Sigma(\sigma) = \mathbf{EDP}_\Sigma(\sigma \cup \mathsf{U})$.*

*3. Given $\sigma_1, \sigma_2 \subseteq \Sigma$, if $\sigma_1 \setminus \sigma_2$ contains*
   *(a) Only unary predicates, then $\mathbf{EDP}_\Sigma(\sigma_2) \subseteq \mathbf{EDP}_\Sigma(\sigma_1)$.*
   *(b) A predicate of arity $\geq 2$, then $\mathbf{EDP}_\Sigma(\sigma_2) \not\subseteq \mathbf{EDP}_\Sigma(\sigma_1)$.*

*Proof*: Statements (1) and (2) above are easy to check and statement (3)(a) follows from (1) and (2). For 3(b), let $P$ be the arity $\geq 2$ predicate. Then consider the sentence $\phi$ as seen in the proof of case 1 of Theorem 2(2) (in section 4.1).

$\phi = (\bigwedge_{Q \in \Sigma \setminus \{P\}} \forall \mathbf{z}_Q Q(\mathbf{z}_Q)) \wedge (\forall x \exists y P(x, y, y, \ldots, y))$

In PCNF, one can check that $\phi \in \mathbf{EDP}_\Sigma(\Sigma \setminus \{P\})$ and hence $\phi \in \mathbf{EDP}_\Sigma(\sigma_2)$. But as shown earlier, for any $\psi$ equivalent to $\phi$, $\psi \notin \mathbf{EBS}_\Sigma(\sigma_1)$ and hence by Corollary 4 (in section 8.1), $\psi \notin \mathbf{EDP}_\Sigma(\sigma_1)$. ∎

Analogous to $\mathcal{EBS}_\Sigma$, we define $\mathcal{EDP}_\Sigma = \{\mathbf{EDP}_\Sigma(\sigma) \mid \sigma \subseteq \Sigma\}$. Then from Theorem 5(2), $\mathcal{EDP}_\Sigma = \{\mathbf{EDP}_\Sigma(\mathsf{U} \cup \sigma) \mid \sigma \subseteq (\Sigma \setminus \mathsf{U})\}$. Then we have the following.

**Corollary 5** $(\mathcal{EDP}_\Sigma, \sqsubseteq)$ *is a lattice which is isomorphic to the powerset lattice $(\wp(\Sigma \setminus \mathsf{U}), \subseteq)$.*

*Proof*: From Theorem 5(3), we know that $\mathbf{EDP}_\Sigma(\mathsf{U} \cup \sigma_1) \neq \mathbf{EDP}_\Sigma(\mathsf{U} \cup \sigma_2)$ for $\sigma_1 \neq \sigma_2$ where $\sigma_1, \sigma_2 \subseteq (\Sigma \setminus \mathsf{U})$. Consider $f : \mathcal{EBS}_\Sigma \to \wp(\Sigma \setminus \mathsf{U})$ such that $f(\mathbf{EDP}_\Sigma(\mathsf{U} \cup \sigma)) = \Sigma \setminus \sigma$. $f$ can be seen to be a bijection.

Now if $f(\mathbf{EDP}_\Sigma(\mathsf{U} \cup \sigma_1)) \subseteq f(\mathbf{EDP}_\Sigma(\mathsf{U} \cup \sigma_2))$, then $\sigma_2 \subseteq \sigma_1$ and hence by parts (1) and (2) of Theorem 5, $\mathbf{EDP}_\Sigma(\mathsf{U} \cup \sigma_1) \subseteq \mathbf{EDP}_\Sigma(\mathsf{U} \cup \sigma_2)$. Conversely if $\mathbf{EDP}_\Sigma(\mathsf{U} \cup \sigma_1) \subseteq \mathbf{EDP}_\Sigma(\mathsf{U} \cup \sigma_2)$, then since $\sigma_1$ and $\sigma_2$ cannot contain any unary predicates, by Theorem 5(3), it must be that $\sigma_2 \setminus \sigma_1$ is empty. In other words, $f(\mathbf{EDP}_\Sigma(\mathsf{U} \cup \sigma_1)) \subseteq f(\mathbf{EDP}_\Sigma(\mathsf{U} \cup \sigma_2))$. Thus $f$ is an isomorphism from $\mathcal{EBS}_\Sigma$ to $\wp(\Sigma \setminus \mathsf{U})$. ∎

The *lub*($\sqcup$) and *glb*($\sqcap$) operators in $(\mathcal{EDP}_\Sigma, \sqsubseteq)$ are defined as: $\mathbf{EDP}_\Sigma(\sigma_1) \sqcup \mathbf{EDP}_\Sigma(\sigma_2) = \mathbf{EDP}_\Sigma(\sigma_1 \cap \sigma_2)$ and $\mathbf{EDP}_\Sigma(\sigma_1) \sqcap \mathbf{EDP}_\Sigma(\sigma_2) = \mathbf{EDP}_\Sigma(\sigma_1 \cup \sigma_2)$. Note that $\mathbf{EDP}_\Sigma(\sigma_1 \cup \sigma_2) \subseteq (\mathbf{EDP}_\Sigma(\sigma_1) \cap \mathbf{EDP}_\Sigma(\sigma_2))$ and $(\mathbf{EDP}_\Sigma(\sigma_1) \cup \mathbf{EDP}_\Sigma(\sigma_2)) \subseteq \mathbf{EDP}_\Sigma(\sigma_1 \cap \sigma_2)$.

Finally the following result shows that except under certain conditions, $\mathbf{EDP}_\Sigma(\sigma)$ is not a syntactic characterization of $\mathbf{EBS}_\Sigma(\sigma)$ in general. Below $A \sqsubset B$ means $A \sqsubseteq B$ but $B \not\sqsubseteq A$. Also $\cong$ means 'semantic equivalence' i.e. $A \cong B$ iff $A \sqsubseteq B$ and $B \sqsubseteq A$.

**Theorem 6** *1. If $\Sigma$ contains only unary predicates, $\mathbf{EDP}_\Sigma(\sigma) \cong \mathbf{EBS}_\Sigma(\sigma) \cong \mathbf{BSR}_\Sigma$ for all $\sigma \subseteq \Sigma$.*

2. If $\Sigma$ contains a predicate of arity $\geq 2$, then
   (a) $\mathbf{EDP}_\Sigma(\Sigma) \cong \mathbf{EBS}_\Sigma(\Sigma) \cong \mathbf{BSR}_\Sigma$.
   (b) If $\Sigma \setminus \sigma$ contains atleast one unary predicate, then $\mathbf{EDP}_\Sigma(\sigma) \sqsubset \mathbf{EBS}_\Sigma(\sigma)$.

*Proof:* Part 1 above follows from Theorem 2(3) (of section 4.1) and Corollary 4 (of section 8.1). Part 2 follows from Theorem 1 (of section 3) and Corollary 4. For part 2, let $U$ be the said unary predicate. Let $P \in \Sigma$ be a predicate of arity $\geq 2$. Then consider the sentence $\phi$ used in the proof of case (2) of Theorem 2(2). Now $\phi \in \mathbf{EBS}_\Sigma(\sigma)$. But suppose it is equivalent to $\psi \in \mathbf{EDP}_\Sigma(\sigma)$. By Theorem 5(2) and Corollary 4, $\psi \in \mathbf{EBS}_\Sigma(\sigma \cup \mathsf{U})$. But then from the proof of case (2) of Theorem 2(2), $\psi$ cannot be in any $\mathbf{EBS}_\Sigma(\sigma_1)$ where $\sigma_1$ contains $\mathsf{U}$. ∎

We do not know whether $\mathbf{EBS}_\Sigma(\sigma)$ strictly semantically subsumes $\mathbf{EDP}_\Sigma(\sigma)$ when $\Sigma \setminus \sigma$ does not contain any unary predicate.

## 8.3 Extensions of the $\mathbf{EDP}_\Sigma(\sigma)$ class

### 8.3.1 Relaxing the last condition of $\mathbf{EDP}_\Sigma(\sigma)$

Consider the last condition in the $\mathbf{EDP}_\Sigma(\sigma)$ definition disjuncted with the following condition:

For a predicate $P$ of arity $\geq 2$, for every pair of distinct instances of $P$, atleast one of which is existential, either (a) the two instances are existentially distinguishable w.r.t. $v \in \overline{E_\mathsf{U}}$ or (b) the two instances are existentially distinguishable w.r.t to every variable in $E_{1,2}$ where $E_{1,2}$ is the set of all the $E_U$ variables appearing as arguments in the two instances. Then the bound in this case is given by $\mathcal{B} = |V(\varphi)| + |\overline{E_\mathsf{U}}| + |E_\mathsf{U}| \cdot 2^k$.

*Proof:*
We refer to the proof of Theorem 4 as given in section 8.1. We will need expand the initial substructure $M_1$ by making only the following changes. We choose **Val**(*Free*) and **Val**($E_\mathsf{U}$) as before. For any colour $c$, $A^c$ however is chosen as follows: Consider the set $\mathcal{V}_M^c$ as defined in the proof. Let $|E_\mathsf{U}| = p$. If $|\mathcal{V}_M^c| < p$, then choose $A^c = \mathcal{V}_M^c$ (in this case we call $A^c$ to be a 'small partition'). Else let $a_{j_1}^c, \ldots, a_{j_p}^c$ be $p$ distinct elements of $\mathcal{V}_M^c$ where $E_\mathsf{U} = \{v_{j_1}, \ldots, v_{j_p}\}$. Then choose $A^c = \{a_{j_1}^c, \ldots, a_{j_p}^c\}$ (in this case we call $A^c$ as 'a subset of a large partition'). As before, let $M_1$ be the substructure generated by **Val**(*Free*), **Val**($\overline{E_\mathsf{U}}$) and $\bigcup_{c \in \{0,1\}^k} A^c$. Then

$$|U_M| \leq |\mathbf{Val}(Free)| + |\mathbf{Val}(\overline{E_\mathsf{U}})| + \Sigma_{c \in \{0,1\}^k}|A^c|$$
$$\leq |V(\varphi)| + |\overline{E_\mathsf{U}}| + |E_\mathsf{U}| \cdot 2^k$$

so that $\mathcal{B} = |V(\varphi)| + |\overline{E_\mathsf{U}}| + |E_\mathsf{U}| \cdot 2^k$.

Let **Z** be an instantiation of the universal variables. We modify *Selector* (which chooses the values to be assigned to $v_i \in E_\mathsf{U}$) as follows. Let $d$ be the value from $M$ chosen for $v_i$.

1. If $d \in \mathbf{Val}(Free)$, this same value is chosen in $M_3$.

2. Else, let $c$ be the colour of $d$ in $M$. We have two subcases here.
   (a) If $d$ is from a small partition, then $d$ itself is chosen as the value for $v_i$
   (b) Else $a_i^c$ is chosen as the value of $v_i$.

Consider an instantiation **Z** of the universal variables. Suppose $P$ satisfies the condition above. Then for such a $P$, we perform the assignment of truth values to the (instantiated) instances of $P$ in $\psi_{M_3}[\mathbf{Z}]$ as follows: for each instance $I$ of $P$ in $\psi$, $I_{M_3}[\mathbf{Z}]$ is assigned the same truth value in $M_3$ as $I_M[\mathbf{Z}]$ is assigned in $M$ (This kind of assignment is only for $P$ which satisfies the above condition. For existential predicates $R$ of arity $\geq 2$ which satisfy condition (3) of the 'basic' **EDP** definition, the assignment is as given in the proof of Theorem 4 of section 8.1).
All that we need to show now is that this assignment does not produce any conflicts for $P$.

Suppose two instances conflicted. Let $I_{M_3}^t[\mathbf{Z}_t]$ and $I_{M_3}^f[\mathbf{Z}_f]$ be the conflicting instances. Then $I_{M_3}^t[\mathbf{Z}_t] = I_{M_3}^f[\mathbf{Z}_f] = P(\mathbf{d})$ (say) and w.l.o.g. assume that $M_3 \models I_{M_3}^t[\mathbf{Z}_t]$ while $M_3 \models \neg I_{M_3}^f[\mathbf{Z}_f]$. If $I^t$ and $I^f$ are free or universal instances in $\psi$, then by Lemma 11 (appearing in the proof of Theorem 4), $I_{M_3}^t[\mathbf{Z}_t] = I_M^t[\mathbf{Z}_t]$ and $I_{M_3}^f[\mathbf{Z}_f] = I_M^f[\mathbf{Z}_f]$. Further by our assignment of truth values above, $M_3 \models I_{M_3}^\alpha[\mathbf{Z}_\alpha]$ iff $M \models I_M^\alpha[\mathbf{Z}_\alpha]$ for $\alpha \in \{t, f\}$. Then this would produce a conflict in $M$ itself which we know is absent.
If atleast one of $I^t$ or $I^f$ is existential, then we have the following cases:

1. $I^t$ and $I^f$ are existentially distinguishable w.r.t. $v \in \overline{E_\mathsf{U}}$. Then in this case a conflict cannot occur by a similar reasoning as in the proof of Theorem 4.

2. For each $v \in \overline{E_\mathsf{U}}$, the $i^{th}$ argument of $I^t$ is $v$ iff the $i^{th}$ argument of $I^f$ is $v$ (i.e. no $v$ existentially *distinguishes* $I^t$ and $I^f$). Then as per the new condition, $I^t$ and $I^f$ are existentially distinguishable w.r.t. all variables in $E_{t,f}$ where $E_{t,f}$ is the set of all $E_\mathsf{U}$ variables that appear as arguments in $I^t$ and $I^f$.
Suppose all the $E_\mathsf{U}$ variables in $I_{M_3}^t[\mathbf{Z}_t]$ and $I_{M_3}^f[\mathbf{Z}_f]$ take values from a small partition. Then from case (2) of *Selector* above, each $E_\mathsf{U}$ variable in $I^t$ takes on the same value in $I_{M_3}^t[\mathbf{Z}_t]$ as in $I_M^t[\mathbf{Z}_t]$. Likewise for $I^f$. If the $i^{th}$ argument of $I^t$ is $v \in \overline{E_\mathsf{U}}$, then so is the $i^{th}$ argument of $I^f$. Putting together these observations with the fact that $I_{M_3}^t[\mathbf{Z}_t] = I_{M_3}^f[\mathbf{Z}_f]$, we can infer that $I^t_M[\mathbf{Z}_t] = I^f_M[\mathbf{Z}_f]$. But since the truth values are preserved for corresponding instances in $\psi_M[\mathbf{Z}]$ and $\psi_{M_3}[\mathbf{Z}]$, we conclude that there must be a conflict in $M$ itself which we know is absent.

The case remaining is that there is some $E_\mathsf{U}$ variable $v_i$ in $E_{t,f}$ that takes on a value from a subset of a large domain i.e. $v_i$ is assigned a value of the form $a^c_i$. Suppose $v_i$ appears as the $g^{th}$ argument of $I^t$. Then the $g^{th}$ argument of $I^t_{M_3}[\mathbf{Z}_t]$ is $a^c_i$. Now since $I^t$ and $I^f$ are existentially distinguishable w.r.t. $v_i$, the $g^{th}$ argument of $I^f$ is either (a) $x \in$ *Free* or (b) $v \in \overline{E_\mathsf{U}}$ or (c) $v_j \in E_\mathsf{U}$ where $j \neq i$. Then the $g^{th}$ argument of $I^f_{M_3}[\mathbf{Z}_f]$ is either (a) a value from **Val**(*Free*) or (b) a value from **Val**($\overline{E_\mathsf{U}}$) or (c) a value from a small partition or a value of the form $a^c_j$ (which is not equal to $a^c_i$ as $j \neq i$. In each of these cases, we can see that the $g^{th}$ argument of $I^f_{M_3}[\mathbf{Z}_f]$ cannot be $a^c_i$. Then $I^t_{M_3}[\mathbf{Z}_t]$ and $I^f_{M_3}[\mathbf{Z}_f]$ do not conflict in the first place.

Thus in each case we have shown that conflicts are not possible at all.

Note that since the colour of the value assigned to $v_i$ in $\psi_M[\mathbf{Z}]$ and $\psi_{M_3}[\mathbf{Z}]$ is the same, the argument given for the unary predicates in the 'Completing the proof' section of the proof of Theorem 4 would still hold. ∎

### 8.3.2 Relaxing the equalities

1. If equalities are also allowed between free and $E_\mathsf{U}$ variables (in addition to the ones given in the **EDP** definition), then the bound remains the same as the one given in Theorem 4).

*Proof:*
We use the same construction as given in the proof of Theorem 4 in section 8.1 and show below that it also handles the case of equalities mentioned above. Consider the case when equalities are also allowed between free and $E_\mathsf{U}$ variables. Suppose the equality $x = v_i$ occurs in some clause where $x \in$ *Free* and $v_i \in E_\mathsf{U}$. Now according to our choice of the values for variable $v_i \in E_\mathsf{U}$ given an instantiation $\mathbf{Z}$ of the universal variables, if in the model $M$, it took on the value in **Val**(*Free*), then in the model $M_3$, we choose the same value for it. Thus the equality $x = v_i$ evaluates to the same truth value in $M_3$ as it does in $M$. If $v_i$ took on a value $d \notin$ **Val**(*Free*), then $x = v_i$ would evaluate to False in $M$. In $M_3$, $v_i$ would be assigned the value $a^c$ where $c$ is the colour of $d$ in $M$. Since $a^c \in A^c$ and $A^c \cap$ **Val**(*Free*) $= \emptyset$, $a^c \notin$ **Val**(*Free*) and hence $x = v_i$ would evaluate to False in $M_3$ as well. Thus in either case, we find that the equality $x = v_i$ evaluates to the same truth value in $M_3$ as it does in $M$. Thus the choice of $v_i$ as done by *Selector* handles the case of equalities between free and $E_\mathsf{U}$ variables as well. ∎

2. If equalities are also allowed between two $E_\mathsf{U}$ variables (in addition to the above equalities), then the bound is given by $\mathcal{B} = |V(\varphi)| + |\overline{E_\mathsf{U}}| + |E_\mathsf{U}| \cdot 2^k$.

*Proof:*
Consider the case when equalities are allowed between two $E_\mathsf{U}$ variables. We will need expand the initial substructure $M_1$ as constructed in the proof of Theorem 4, by making only the following changes. We choose **Val**(*Free*) and **Val**($E_\mathsf{U}$) as before. For any colour $c$, $A^c$ however is chosen as follows: Consider the set $\mathcal{V}^c_M$ as defined earlier. Let $|E_\mathsf{U}| = p$. If $|\mathcal{V}^c_M| < p$, then choose $A^c = \mathcal{V}^c_M$. Else let $a^c_{j_1}, \ldots, a^c_{j_p}$ be $p$ distinct elements of $\mathcal{V}^c_M$ where $E_\mathsf{U} = \{v_{j_1}, \ldots, v_{j_p}\}$. Then choose $A^c = \{a^c_{j_1}, \ldots, a^c_{j_p}\}$. As before, let $M_1$ be the substructure generated by **Val**(*Free*), **Val**($\overline{E_\mathsf{U}}$) and $\bigcup_{c \in \{0,1\}^k} A^c$. Then

$$\begin{aligned}|U_M| &\leq |\mathbf{Val}(\textit{Free})| + |\mathbf{Val}(\overline{E_\mathsf{U}})| + \Sigma_{c \in \{0,1\}^k}|A^c| \\ &\leq |V(\varphi)| + |\overline{E_\mathsf{U}}| + |E_\mathsf{U}| \cdot 2^k\end{aligned}$$

so that $\mathcal{B} = |V(\varphi)| + |\overline{E_\mathsf{U}}| + |E_\mathsf{U}| \cdot 2^k$.

Let $\mathbf{Z}$ be an instantiation of the universal variables. We modify *Selector* (which chooses the values to be assigned to $v_i \in E_\mathsf{U}$) as follows. Let $d$ be the value from $M$ chosen for $v_i$.

1. If $d \in$ **Val**(*Free*), this same value is chosen in $M_3$.

2. Else, let $c$ be the colour of $d$ in $M$. We have two subcases here.
   (a) If $v_i$ is assigned a value which is different from the values assigned to the other $v_j$'s where $v_j \in E_\mathsf{U}$, then choose $a^c_i$ for $v_i$ in $M_3$.
   (b) Let $V$ be the set of all those $v_j$'s which have been assigned the same value as $v_i$. Then let $i^*$ be the least index appearing as subscript of some variable in $V$. Then choose $a^c_{i^*}$ for $v_i$ in $M_3$.

We will now show that by the choices made as above, the new kinds of equalities in $\psi_{M_3}[\mathbf{Z}]$ will also evaluate to the same truth value in $M_3$ as the corresponding equalities in $\psi_M[\mathbf{Z}]$ evaluate to in $M$.

Suppose the equality $x = v_i$ occurs in some clause where $x \in$ *Free* and $v_i \in E_\mathsf{U}$. Now according to our choice of the values for variable $v_i \in E_\mathsf{U}$ given an instantiation $\mathbf{Z}$ of the universal variables, if in $\psi_M[\mathbf{Z}]$, it took on the value in **Val**(*Free*), then in the $\psi_{M_3}[\mathbf{Z}]$ also the same value is chosen for it. Thus the equality $x = v_i$ evaluates to the same truth value in $M_3$ as it does in $M$. If $v_i$ took on a value $d \notin$ **Val**(*Free*), then $x = v_i$ would evaluate to False in $M$. In $M_3$, $v_i$ would be assigned a value in $A^c$ which is disjoint with **Val**(*Free*). Hence $x = v_i$ would evaluate to False in $M_3$ as well. Thus in either case, we find that the equality $x = v_i$ evaluates to the same truth value in $M_3$ as it does in $M$.

Suppose the equality $v_i = v_j$ occurs in some clause where $v_i, v_j \in E_\cup$. Suppose in $\psi_M[\mathbf{Z}]$, $v_i$ and $v_j$ are assigned the same value. Then from case 2(b) of the modified *Selector* above, we know that in $\psi_{M_3}[\mathbf{Z}]$ too, $v_i$ and $v_j$ would be assigned the same value. If $v_i$ and $v_j$ were assigned different values in $\psi_M[\mathbf{Z}]$, then they would be assigned different values in $\psi_{M_3}[\mathbf{Z}]$ as well. If not then say $a_k^c$ is the common value assigned to both in $\psi_{M_3}[\mathbf{Z}]$. Then from case 2(b) of *Selector*, it must have been the case that $v_i$ was assigned the same value as $v_k$ (where $v_k$ is the least indexed variable in the set of all the $E_\cup$ variables whose values in $\psi_M[\mathbf{Z}]$ were the same as $v_i$) and also $v_j$ was assigned the same value as $v_k$ in $\psi_M[\mathbf{Z}]$. Then $v_i$ and $v_j$ were assigned the same value which contradicts the assumption made above. Thus in either case $v_i = v_j$ evaluates to the same truth value in $M_3$ as it does in $M$.

Note that since the colour of the value assigned to $v_i$ in $\psi_M[\mathbf{Z}]$ and $\psi_{M_3}[\mathbf{Z}]$ is the same, the argument given for the unary predicates in the 'Completing the proof' section of the proof of Theorem 4 would still hold. ∎

*Conjecture:*

Generalising **EDP** with the conditions 1 and 2(b) above together, the bound is given by

$$\mathcal{B} = |V(\varphi)| + |\overline{E_\cup}| + |E_\cup| \cdot 2^k$$

#### 8.3.3 Extension to an order-sorted logic

There has been recent work ([8]) on extending **BSR** to order-sorted logics. $\mathbf{EDP}_\Sigma(\sigma)$ being an unsorted logic which is orthogonal to the extension presented in [8], these two extensions can be combined to yield syntactic generalizations of **BSR** which subsume each of the individual extensions.

### 8.4 The Löwenheim class with equality

For the Löwenheim class with equality, the bound $\mathcal{B}$ in general is $q.2^k$ (where $q$ is the length of the quantifier prefix) [2]. But this class, being a special case of the $\mathbf{EDP}_\Sigma(\Sigma)$ class, from the above results, we get the following finer bounds for the Löwenheim class with equality for different cases of equality.

1. For the Löwenheim class extended with equality, where the equalities are either between (a) two free variables or (b) two universal variables or (c) a free and a universal variable or (d) a free and an $E_\cup$ variable, the bound (on the size of the sub-model) is given by $\mathcal{B} = |V(\varphi)| + 2^k$.

2. For the Löwenheim class extended with equality, if in addition to the above kind of equalities, equalities between $E_\cup$ variables are also allowed, then the bound is given by $\mathcal{B} = |V(\varphi)| + |E_\cup| \cdot 2^k$.

### 8.5 Complexity of the SAT problem for EDP

The bounded model property of $\mathbf{EDP}_\Sigma(\sigma)$ allows us to also characterize the complexity of its SAT problem.

**Theorem 7** *The satisfiability problem for $\mathbf{EDP}_\Sigma(\sigma)$ is* NEXPTIME-*complete.*

*Proof:* Indeed $\mathbf{EDP}_\Sigma(\sigma)$ satisfies the two conditions of Corollary 2 (appearing towards the end of section 5.3): (a) The bound $\mathcal{B}$ for a sentence $\phi \in \mathbf{EDP}_\Sigma(\sigma)$ is $\leq 2^{O(|\phi|)}$. Then $f(\phi)$, as defined in Lemma 8 (of section 5.3), is efficiently (polynomial time) computable and such that $f(\phi) \leq 2^{O(|\phi|)}$. (b) $\mathbf{EDP}_\Sigma(\sigma)$ contains $\mathbf{BSR}_\Sigma$ (since $\mathbf{BSR}_\Sigma \subseteq \mathbf{EDP}_\Sigma(\Sigma) \subseteq \mathbf{EDP}_\Sigma(\sigma)$). ∎

An interesting consequence of the above result is that since the satisfiability checking problem for both $\mathbf{BSR}_\Sigma$ and $\mathbf{EDP}_\Sigma(\sigma)$ is NEXPTIME-complete for all $\sigma \subseteq \Sigma$, it means that there is an efficient reduction from $\mathbf{EDP}_\Sigma(\emptyset)$ to $\mathbf{BSR}_\Sigma$ which yields a $\mathbf{BSR}_\Sigma$ sentence which is equisatisfiable with a given $\mathbf{EDP}_\Sigma(\emptyset)$ sentence. There is some recent work ( [3]) on improved heuristics in deciding $\mathbf{BSR}_\Sigma$ sentences. Finding an efficient reduction from $\mathbf{EDP}_\Sigma(\emptyset)$ to $\mathbf{BSR}_\Sigma$ would therefore help us get better heuristics to decide $\mathbf{EDP}_\Sigma(\emptyset)$ sentences as well. We present one such reduction in Section 10.

## 9 Closure Properties of $\mathcal{EBS}_\Sigma$ and $\mathcal{EDP}_\Sigma$

For $\Sigma$ which contains only unary predicates, $\mathbf{EBS}_\Sigma(\sigma) = \mathbf{EDP}_\Sigma(\sigma) = \mathbf{FO}_\Sigma$. Clearly then $\mathbf{EBS}_\Sigma(\sigma) = \mathbf{EDP}_\Sigma(\sigma)$ are closed under $\wedge, \vee$ and $\neg$.
Hence below we consider $\Sigma$ to have atleast one arity $\geq 2$ predicate.

### 9.1 Closure properties of $\mathcal{EBS}_\Sigma$

Consider $\varphi_i \in \mathbf{EBS}_\Sigma(\sigma_i)$. Let $\mathcal{B}_1, \mathcal{B}_2$ be the bounds for $\varphi_1, \varphi_2$ respectively.

## 1) ∧−closure

Let $\varphi = \varphi_1 \wedge \varphi_2$. Assume $\sigma_1 = \sigma_2 = \Sigma$.
Suppose $M \models \varphi$. Then $M \models \varphi_i$. Then there exists a substructure $M_{i,1}$ of $M$ such that $|\mathcal{U}_{M_{i,1}}| \leq \mathcal{B}_i$ which satisfies the $\mathbf{EBS}_\Sigma(\Sigma)$ conditions for $\varphi_i$. Then let $M_1$ be the substructure of $M$ generated by $\mathcal{U}_{M_{1,1}} \cup \mathcal{U}_{M_{2,1}}$. Consider any extension $M_2$ of $M_1$ within $M$. Since $M_2$ extends $M_{i,1}$, $M_2 \models \varphi_i$ (this is because $\varphi_i \in \mathbf{EBS}_\Sigma(\sigma_i)$). Then $M_2 \models \varphi$. Thus $\varphi \in \mathbf{EBS}_\Sigma(\Sigma)$. The bound $\mathcal{B}$ of $\varphi$ is not more than $|U_{M_1}|$ which is $\leq \mathcal{B}_1 + \mathcal{B}_2$.
However suppose w.l.o.g, $\sigma_1$ is not $\Sigma$. Then while considering any common extension $M_2$ of $M_{1,1}$ and $M_{2,1}$ (mentioned above), to satisfy $\varphi_1$, in general, we might have to interpret the $\Sigma \setminus \sigma_1$ predicates in a way which might conflict with the way these predicates are interpreted in trying to satisfy $\varphi_2$.

## 2) ∨−closure

Let $\varphi = \varphi_1 \vee \varphi_2$.
Suppose $M \models \varphi$. Then suppose $M \models \varphi_1 \wedge \neg \varphi_2$. Then consider $M_1$ as $M_{1,1}$ mentioned above. Then any extension $M_2$ of $M_1$ within $M$ satisfies $\varphi_1$ and hence $\varphi$. If $M \models \varphi_2 \wedge \neg \varphi_1$ then consider $M_1$ as $M_{2,1}$ mentioned above. Then any extension $M_2$ of $M_1$ within $M$ satisfies $\varphi_2$ and hence $\varphi$. Thus in either case $M_1$ is such that $|U_{M_1}| \leq \max\{\mathcal{B}_1, \mathcal{B}_2\}$. Further, the interpretations of the $\sigma_1 \cap \sigma_2$ are guaranteed to be preserved when extending $M_{1,1}$ or $M_{2,1}$. Hence $\varphi \in \mathbf{EBS}_\Sigma(\sigma_1 \cap \sigma_2)$. The bound for $\varphi$ is $\mathcal{B} = |U_{M_1}| \leq \max\{\mathcal{B}_1, \mathcal{B}_2\}$.

## 3) ¬−(non)closure

Consider the following sentence in the vocabulary $\Sigma = \{E\}$ where $E$ is a 2-ary predicate (If $\Sigma$ contains a higher arity predicate, then that can simulate a 2-ary predicate and hence we can construct an example in that case too similar to the one shown below).

$$\begin{aligned}\Phi &= \forall x \neg E(x,x) \wedge \forall x \exists y\, ((x \neq y) \wedge E(x,y)) \wedge \\ &\quad \forall x \forall y \forall z (E(x,y) \wedge E(y,z) \to E(x,z)) \\ &= \forall x \forall y \forall z \exists w \\ &\quad (\neg E(x,x) \wedge (E(x,y) \wedge E(y,z) \to E(x,z))\wedge \\ &\quad (x \neq w) \wedge E(x,w))\end{aligned}$$

We can see that the models of this sentence are infinite DAGs. Then $\Phi \notin \mathbf{EBS}_\Sigma(\emptyset)$. Now $\Phi$ is a $\forall^*\exists^*$ sentence and its negation is an $\exists^*\forall^*$ sentence which belongs to $\mathbf{BSR}_\Sigma \subseteq \mathbf{EBS}_\Sigma(\Sigma)$.

## 9.2 Closure properties of $\mathcal{EDP}_\Sigma$

Consider $\varphi = \varphi_1 \wedge \varphi_2$ where $\varphi_i \in \mathbf{EDP}_\Sigma(\sigma_i)$.

## 1) ∧−closure

Let $\varphi = \varphi_1 \wedge \varphi_2$.
Let $\sigma_i^*$ be the set of all unary, free and universal predicates appearing in $\varphi_i$. Assume $\sigma_1^* = \sigma_2^* = \Sigma$. Then there are no existential predicates in both $\varphi_1$ and $\varphi_2$. Then in $\varphi$, firstly rename the variables of $\varphi_2$ so that they are completely disjoint with those of $\varphi_1$ and bring $\varphi$ in PCNF with the leftmost $\exists$ quantifiers of $\varphi$ being leftmost $\exists$ quantifiers of $\varphi_1$ and $\varphi_2$ and the matrix of $\varphi$ being the matrices of $\varphi_1$ and $\varphi_2$ conjuncted. Then $V(\varphi) = V(\varphi_1) \cup V(\varphi_2)$. The predicates which were free or universal in $\varphi_1$ or $\varphi_2$ continue to remain so in $\varphi$ since the free or universal nature of each instance of $\varphi_1$ or $\varphi_2$ remains unaffected upon the conjunction since we made the variables of $\varphi_1$ and $\varphi_2$ disjoint. Thus $\mathcal{F} = \mathcal{F}_1 \cup \mathcal{F}_2$ and $\mathcal{A} = \mathcal{A}_1 \cup \mathcal{A}_2$. Also clearly $\mathsf{U} = \mathsf{U}_1 \cup \mathsf{U}_2$ i.e. the unary predicates $\mathsf{U}$ of $\varphi$ are those that appear in $\varphi_1$ (i.e. $\mathsf{U}_1$) or those that appear in $\varphi_2$ (i.e. $\mathsf{U}_2$). Then $\Sigma = \sigma_1^* = \sigma_2^* = \sigma_1^* \cup \sigma_2^* = \mathsf{U} \cup \mathcal{F} \cup \mathcal{A} \subseteq \Sigma$. Since equality $\in \mathcal{F}_i \cup \mathcal{A}_i$, we have that in $\varphi$, equality $\in \mathcal{F} \cup \mathcal{A}$. Since there are no existential predicates in $\varphi_i$, there are none in $\varphi$ as well. Thus $\varphi \in \mathbf{EDP}_\Sigma(\Sigma)$.
However suppose w.l.o.g $\sigma_1^* \neq \Sigma$. Then consider the following sentences: $\varphi_1 = \forall x \exists y P(x,y)$ , $\varphi_2 = \forall x \neg P(x,x)$. Then bringing $\varphi$ into PCNF as explained above, we have $\varphi = \forall z \forall x \exists y (P(x,y) \wedge \neg P(z,z))$. These two instances are not existentially distinguishable w.r.t. $y$ - the only variable in $\overline{E_\mathsf{U}}$. Then $\varphi \notin \mathbf{EDP}_\Sigma(\sigma)$ for any $\sigma \subseteq \Sigma$. (One can construct a similar example for any other $\Sigma$ which contains atleast one arity $\geq 2$ predicate.)

## 2) ∨−closure

Let $\varphi = \varphi_1 \vee \varphi_2$.
Assume $\sigma_1^* = \sigma_2^* = \Sigma$. Then again there are no existential variables at all. Then by the same argument as above, it is easy to see that $\varphi \in \mathbf{EDP}_\Sigma(\Sigma)$.
However suppose w.l.o.g $\sigma_1^* \neq \Sigma$. Then consider the following sentences:

$$\varphi_1 = \forall x \exists y (P(x,y) \wedge P(x,x))\ ;\ \varphi_2 = \forall x \neg P(x,x).$$

Then bringing $\varphi$ into PCNF as explained above, we have $\varphi = \forall z \forall x \exists y (P(x,y) \vee \neg P(z,z)) \wedge (P(x,x) \wedge \neg P(z,z))$. Now $P(x,y)$ and $\neg P(z,z)$ appearing in the first and second clauses respectively differ in polarity. But these two instances are not existentially distinguishable w.r.t. $y$ - the only variable in $\overline{E_\mathsf{U}}$. Then $\varphi \notin \mathbf{EDP}_\Sigma(\sigma)$ for any $\sigma \subseteq \Sigma$. (One can construct a similar example for any other $\Sigma$ which contains atleast one arity $\geq 2$ predicate.)

## 3) ¬−(non)closure

Consider $\varphi = \exists x \forall y (P(x,y) \vee \neg P(y,y))$. Now $\varphi \in \mathbf{EDP}_\Sigma(\Sigma)$. But in $\neg \varphi = \forall x \exists y (\neg P(x,y)) \wedge (P(y,y))$, the two instances are not existentially distinguishable w.r.t. $y$ - the only variable in $\overline{E_\mathsf{U}}$. Then $\varphi \notin \mathbf{EDP}_\Sigma(\sigma)$ for any $\sigma \subseteq \Sigma$. (One can construct a similar example for any other $\Sigma$ which contains atleast one arity $\geq 2$ predicate.)

## 10 Spectral Properties of $\mathbf{EDP}_\Sigma(\emptyset)$ and Applications

Consider an $\mathbf{EDP}_\Sigma(\emptyset)$ sentence

$$\varphi = \forall \mathbf{z}_0 \exists v_1 \forall \mathbf{z}_1 \exists v_2 \forall \mathbf{z}_2 \ldots \exists v_r \forall \mathbf{z}_r \xi(\mathbf{z}, v_1 \ldots, v_r)$$

where $\mathbf{z} = (\mathbf{z}_0, \ldots, \mathbf{z}_r)$ and $\xi$ is quantifier-free. Let $\mathcal{B}$ be the bound for $\varphi$ as given by Theorem 4 (of section 8.1), and let $\mathcal{X} = \{x_1, x_2, \ldots x_\mathcal{B}\}$ be a set of $\mathcal{B}$ fresh variables. In addition, let $\psi$ be the $\mathbf{BSR}_\Sigma$ formula defined as follows.
$\psi = \exists x_1 \ldots \exists x_\mathcal{B} \forall \mathbf{z}_0 \forall \mathbf{z}_1 \ldots \forall \mathbf{z}_r (\chi)$ where
$\chi = (\bigvee_{u_1 \in \mathcal{X}} \ldots \bigvee_{u_r \in \mathcal{X}} \xi(\mathbf{z}, v_1 \mapsto u_1, \ldots, v_r \mapsto u_r))$
where $\xi(\mathbf{z}, v_1 \mapsto u_1, \ldots, v_r \mapsto u_r)$ is the formula obtained by replacing every occurence of $v_i$ in $\xi$ by $u_i$.

**Theorem 8** *The spectrum of $\varphi$ equals the spectrum of $\psi$.*

*Proof Sketch:* It is straightforward to see that every model of $\psi$ is also a model of $\varphi$. Therefore, the spectrum of $\psi$ is contained in that of $\varphi$.
For the other way round, consider a model $M$ of $\varphi$. The technique used in the proof of Theorem 4 of section 8.1 can now be used to construct a model $M'$ of $\varphi$ such that $\mathcal{U}_M = \mathcal{U}_{M'}$ and $M \mid_\sigma = M' \mid_\sigma$. Furthermore, the construction guarantees that when $\varphi$ is evaluated on $M'$, all existentially quantified variables can take values from a fixed subset $V \subset \mathcal{U}_{M'}$ such that $|V| \leq \mathcal{B}$, and yet cause $\varphi$ to evaluate to True. This special property of $M'$ ensures that $M' \models \psi$ as well. Since $\mathcal{U}_M = \mathcal{U}_{M'}$, this proves that the spectrum of $\varphi$ is contained in the spectrum of $\psi$. Hence, $\varphi$ and $\psi$ have the same spectrum. Note that $\psi$ is not semantically equivalent to $\varphi$. It is certainly equisatisfiable to $\varphi$, but is also "equispectral" to $\varphi$. ∎

An immediate application of Theorem 8 in SMT solving stems from recent results due to Fontaine [4]. In [4], a notion of *gentle* theories is introduced and it is argued that **BSR**, Löwenheim and $FO^2$ (**FO** with two variables) theories are gentle. Intuitively, a gentle theory is one in which one can completely express the spectrum. Fontaine further shows that gentle theories can be combined with almost any other decidable theories in an SMT solving setting with cooperating decision procedures for multiple theories. From Theorem 8, we know that for every $\varphi \in \mathbf{EDP}_\Sigma(\emptyset)$, we can effectively construct a **BSR** formula $\psi$ that has the same spectrum as $\varphi$. Since **BSR** has been shown to be a gentle theory in [4], this implies that $\mathbf{EDP}_\Sigma(\emptyset)$ is gentle as well, and can be combined with other decidable theories for SMT solving. From Theorem 5 (in section 8.2) and the fact that $\mathbf{EDP}_\Sigma(\Sigma)$ contains $\mathbf{BSR}_\Sigma$, we know that $\mathbf{EDP}_\Sigma(\emptyset)$ semantically generalizes $\mathbf{BSR}_\Sigma$. Hence, $\mathbf{EDP}_\Sigma(\emptyset)$ is a semantically richer theory than $\mathbf{BSR}_\Sigma$ that is still gentle. Interestingly, Fontaine gives separate proofs of gentleness of the **BSR** class and of the Löwenheim class with equality [4]. Since Löweheim class with equality is contained in $\mathbf{EBS}_\Sigma(\Sigma)$ with the bound $\mathcal{B} = q.2^k$ ($q$ = number of quantifiers, and $k$ = number of distinct monadic predicates), Theorem 1 (of section 3) and its proof allow us to effectively construct a **BSR** formula that is semantically equivalent to a Löwenheim formula with equality. Hence a gentleness proof of **BSR** suffices to prove that the Löwenheim class with equality is also gentle. Section 8.4 also describes how $\mathbf{EDP}_\Sigma(\Sigma)$ (and hence $\mathbf{EDP}_\Sigma(\emptyset)$) contains and indeed generalizes the Löwenheim class with equality. We, therefore, believe that $\mathbf{EDP}_\Sigma(\emptyset)$ and its variants truly expand the set of gentle theories, and permit SMT solving with cooperating decision procedures on richer problems. Note that satisfiability checking for formulas in $\mathbf{EDP}_\Sigma(\emptyset)$ can be reduced to the satisfiability checking problem of the equispectral $\mathbf{BSR}_\Sigma$ formula. Model-finding of $\mathbf{EDP}_\Sigma(\emptyset)$ formulas can be done in the same way as is done for $\mathbf{BSR}_\Sigma$ formulas [1].

Yet other interesting applications of the $\mathbf{EDP}_\Sigma(\emptyset)$ class arise in bounded model checking and inductive property checking. Consider a (possibly infinite) state transition system whose transition relation is given by a first order logic formula $T(s, s')$ on current and next state variables $s$ and $s'$ respectively. Let $I(s)$ and $P(s)$ be predicates on state variables denoting the initial set of states and the set of states with property $P$. In the classical bounded model checking problem, we wish to find if there exists a run of the system for $k$ steps, starting from an initial state and ending in a state satisfying $P$. This can be posed as the satisfiability checking problem of the sentence:

$\varphi_{bmc} = \exists s_0 \ldots \exists s_k \ I(s_0) \wedge T(s_0, s_1) \wedge \ldots T(s_{k-1}, s_k) \wedge P(s_k)$

Similarly, in inductive property checking, we wish to find if there exists a run of the system for $k$ steps, starting from a state satisfying $P$ such that $P$ is violated for the first time in the $k^{th}$ step. This is equivalent to checking the satisfiability of the sentence:

$\varphi_{ind} = \exists s_0 \ldots \exists s_k \ P(s_0) \wedge T(s_0, s_1) \wedge P(s_1) \ldots \wedge$
$\qquad\qquad P(s_{k-1}) \wedge T(s_{k-1}, s_k) \wedge \neg P(s_k)$

It can be shown that if $I(s)$, $T(s, s')$, $P(s)$ and $\neg P(s)$ are in $\mathbf{EDP}_\Sigma(\emptyset)$, then one can obtain the size of a bounded model for both $\varphi_{bmc}$ and $\varphi_{ind}$ from the bounds $\mathcal{B}$ for $I(s)$, $T(s, s')$, $P(s)$ and $\neg P(s)$. The "extensible" bounded submodel property of $\mathbf{EDP}_\Sigma(\emptyset)$ formulae ensures that the bound of the model size for $\varphi_{bmc}$ and $\varphi_{ind}$ grows slowly (linearly) with $k$. The reader is referred to [9] for further details.

## 11 Conclusion

In this paper, we presented a semantic and syntactic generalization of the widely used **BSR** class of first order logic sentences. Our study showed a lattice-structured generalization both in the semantic and syntactic domains with strict inclusions between classes of formulae. The semantic generalization subsumes all $\mathbf{FO}_\Sigma$ formulae with finite and co-finite spectra in the limit. We also showed that several well known decidable classes of $\mathbf{FO}_\Sigma$ sentences are subsumed by our generalization. This gives rise to alternative proofs of some interesting results. The syntactic generalization enjoys special spectral properties, that can be effectively used in applications like SMT solving, bounded model checking and inductive property checking. There are several open questions that still remain. Most important among them is the question of whether a syntactic characterization of $\mathbf{EBS}_\Sigma(\emptyset)$ is possible in the same sense that $\mathbf{BSR}_\Sigma$ syntactically characterizes $\mathbf{EBS}_\Sigma(\Sigma)$. Identifying other useful fragments for which membership is decidable, and the bound $\mathcal{B}$ is efficiently computable presents yet another major challenge for future work.

## References

[1] P. Baumgartner, A. Fuchs, H. de Nivelle, and C. Tinelli. Computing finite models by reduction to function-free clause logic. *Journal of Applied Logic*, 7(1):58–74, March 2009.
[2] E. Börger, E. Grädel, and Y. Gurevich. *The Classical Decision Problem*. Perspectives in Mathematical Logic. Springer, 1997.
[3] L. M. de Moura and N. Bjørner. Deciding effectively propositional logic using dpll and substitution sets. In *IJCAR*, pages 410–425, 2008.
[4] P. Fontaine. Combinations of theories for decidable fragments of first-order logic. In *FroCos*, pages 263–278, 2009.
[5] E. Grädel, P. G. Kolaitis, and M. Y. Vardi. On the decision problem for two-variable first-order logic. *The Bulletin of Symbolic Logic*, 3(1):53–69, 1997.
[6] L. Libkin. *Elements of Finite Model Theory*. ETACS: Texts in Theoretical Computer Science. Springer, 2004.
[7] M. Margenstern. Decidability and undecidability of the halting problem on turing machines, a survey. In *LFCS '97*, pages 226–236, London, UK, 1997. Springer-Verlag.
[8] T. Nelson, D. J. Dougherty, K. Fisler, and S. Krishnamurthi. On the finite model property in order-sorted logic. Technical report, Worcester Polytechnic Institute, (http://web.cs.wpi.edu/~tn/publications/finite-model-osl-TR.pdf).
[9] A. Sankaran. Reachability analysis in graph transformation systems. Technical report, Dept. of Computer Science and Engineering, IIT Bombay, Mumbai (http://www.cse.iitb.ac.in/~abhisekh/reachability-GTS.pdf), 2007.